\documentclass{aa}  

\usepackage{graphicx}

\usepackage{txfonts}

\usepackage{tikz}

\usepackage[colorlinks=true,linkcolor=blue,filecolor=blue,urlcolor=blue,citecolor=blue]{hyperref}

\AtBeginDocument{\mathcode`v=\varv}

\begin{document}

	\title{Full non-LTE multi-level radiative transfer}
	
	\subtitle{II. The case of a 5-level Ca\,{\sc ii} atom with broadened excited levels}
	
	\author{T. Lagache
		\inst{1}    
		\fnmsep\thanks{\email{tristan.lagache@utoulouse.fr}}  
		\and       
		M. Sampoorna
		\inst{2}
		\and
		F. Paletou
		\inst{1} 
	}
	
	\institute{Université de Toulouse, Observatoire Midi-Pyrénées,
          Cnrs, Cnes, Irap, Toulouse, France
          \and Indian Institute of
          Astrophysics, Koramangala, Bengaluru 560034, India
        }
       \date{Received 2 April 2026 / Accepted 12 May 2026}
	
	\abstract
	{The so-called full non-local thermodynamic equilibrium (FNLTE) radiative transfer problem allows us to take into account not only deviations of the radiation field from the Planckian but also deviations of the densities and velocity distributions of massive particles from Maxwell-Boltzmann statistics. This article discusses the extension of this formalism to physically realistic multi-level atoms, including natural broadening of the excited levels. In practice, we must solve self-consistently a coupled set of kinetic equations and determine, for each line, an emission and absorption profile by convolving a non-Lorentzian atomic profile with a non-Maxwellian velocity distribution at each iteration. To solve this numerically challenging problem, we have developed a new efficient iterative method based on well-known approximate operator techniques. After validating our numerical strategy, we present the results obtained for the H\,\&\,K lines and the infrared triplet of the Ca\,{\sc ii}. Under the conditions studied, for this particular atomic model and for a simplified atmosphere, we find that the standard NLTE with partial redistribution is sufficient to describe the formation of Ca\,{\sc ii} spectral lines. The more exact treatment of FNLTE is unnecessary in the case of Ca\,{\sc ii} H\,\&\,K, and infrared triplet lines, even when accounting for velocity-changing collisions.}	
	
	
	\maketitle
	
	\section{Introduction}
	\label{sec:intro}    
		 \defcitealias{LPS25}{LPS25}
		 \defcitealias{NIST_ASD}{NIST}
		 \defcitealias{Gaia}{GAIA}
		 
		In a previous article (\citealt{LPS25}, hereafter LPS25), we extended the so-called full non-local thermodynamic equilibrium (hereafter FNLTE) radiative transfer problem, originally formulated by \citet[see also \citealt{HOSI}, \citealt{Oxenius65} and \citealt{OxeniusSimoneau94}]{OxeniusBook}, to multi-level atoms. In essence, the FNLTE description of radiative transfer must take into account not only deviations of the radiation field from the Planck function, but also deviations of the densities and velocity distributions of massive particles from Maxwell-Boltzmann statistics. This is therefore fundamentally a \textit{multi-distribution} problem that can only be solved by obtaining self-consistently all the non-equilibrium distributions associated with each set of particles present in the medium under study, i.e. by solving a coupled set of kinetic equations containing as many radiative transfer equations (RTEs) as there are radiatively allowed transitions and as many kinetic equilibrium equations (KEEs) as there are populations of massive particles.
		 
		To address this complex problem, it is convenient to first simplify and then, deal with progressively more realistic and complex physical situation. 
		The first solutions to the FNLTE radiation transfer problem were thus obtained for infinitely sharp two-level atoms by \cite{PP21,PSP23,UBU} and then, for a more realistic atomic model with natural broadening of the upper level by \cite{SPV24}. However, this work demonstrated that in the context of two-level atoms, our formalism is completely equivalent to the ``standard'' formalism, which treat only partially the deviations of the massive particles distributions from their equilibrium. Thus, although it constitutes a methodologically necessary step, the specific study of two-level atoms does not lead to the emergence of new behaviours specific to the FNLTE formalism.
		 
		Beyond the simple two-level atom, a multi-level atom problem requires to take into account all transitions and series of transitions populating a given atomic level from lower or higher levels. The coupling of kinetic equations then becomes more important and involved. Each density, each velocity distribution and, consequently, each photon distribution become mutually dependent. The numerical strategies developed up to now, adapted to the self-consistent calculation of only two distributions (one for photons and one for massive particles), must be revised. There is therefore a conceptual and technical barrier between two-level and multi-level atoms cases.
		
		This barrier was first broken by \citetalias{LPS25} in the context of a heuristic atomic model, where all the levels are infinitely sharp. Fundamentally, despite being developed within the framework of a simple atomic model, the tools available to us since \citetalias{LPS25} remain sufficiently general to be adapted to potentially any FNLTE radiative transfer problem. In this article, we propose to study physically realistic atomic models, including the natural broadening of atomic levels, which is essential for understanding the formation of spectral lines. However, including such a mechanism is not without consequences. First, we must determine at each iteration and depth a set of quasi-Voigt emission and absorption profiles, i.e. compute the convolution between non-Lorentzian atomic profiles (instead of a simple Dirac distribution in \citetalias{LPS25}) and non-Maxwellian velocity distribution functions (VDFs). Computing Voigt profiles efficiently and accurately is a problem in itself (see e.g. \citealt{AGQ} and references therein). It is even more complex to do so without prior knowledge of the VDFs and the atomic profiles. Also, the numerical scheme used in \citetalias{LPS25} mainly consists of updating successively all the relevant radiative and kinetic quantities in somewhat close to a $\Lambda$-iteration, which is known to suffer from a convergence problem (see e.g. Chapter 12 of \citealt{HMbook}). Because in the case of infinitely sharp levels studied in \citetalias{LPS25}, we know that the emergent intensity profiles diverge only slightly from the solution obtained in the approximation of the complete redistribution, we initialise the iterative process with this solution in order to achieve convergence. However, by accounting for the broadening of atomic levels, we expect to observe significant deviations from this solution. The \citetalias{LPS25} method is then no longer relevant, and a new numerically efficient iterative method must be developed.
		 
		This article is organised as follows. In Sect.~\ref{sec:Equations}, we present the equations governing potentially any multi-level FNLTE problem, and in Sect.~\ref{sec:MALIBU}, we present a \textit{new} numerical method for solving this problem, which is based on well-known approximate operator techniques. In Sect.~\ref{sec:CaII_model}, we apply the FNLTE formalism to the case of the H\,\&\,K lines and infrared triplet of the Ca\,{\sc ii}, widely used in astrophysics as, respectively, good proxies for stellar activity and stellar parameters. The results for this atomic model are presented in Sect.~\ref{sec:Results}. Finally, in Sect.~\ref{sec:Discussion}, we discuss these new solutions and the future developments needed to include even more realism in our FNLTE formalism.
 
	\section{Statement of the multi-level FNLTE problem}
	\label{sec:Equations}  
	
	In \citetalias{LPS25}, we presented the problem of multi-level FNLTE radiative transfer in its generic form. Regardless of the type of the atomic model we are dealing with, the basic structure of the problem remains unchanged and the equations presented in Sect.~2 of \citetalias{LPS25} remain entirely relevant. We will only recall the essential elements and, for more details, we strongly encourage the reader to refer to \citetalias{LPS25}.
	
	\subsection{Kinetic equations of photons and massive particles}
	
	Hereafter, we consider a model atmosphere composed of a single atomic species interacting with a radiation field. However, the interaction of electrons with photons will not be treated. We fix an electron density and electrons will follow Maxwellian VDF even though, strictly speaking, we should also solve a kinetic equation for them.
	
	For a time-independent problem, without polarisation, photoionisation, continuum, and neglecting stimulated emission, the kinetic equation for photons or RTE is written in 1D plane-parallel geometry as\footnote{For the sake of notational simplicity and brevity, here we modify or simplify some of the notations used in LPS25. Consequently, reader may find some differences in the notations used in this article and LPS25.}: 
	\begin{equation}
		\mu \frac{\partial I_{ij}(x,\mu,\tau)}{\partial \tau} =
		I_{ij}(x,\mu,\tau)  -
		\frac{n_j(\tau) A_{ji}}{n_i(\tau) B_{ij}} \frac{\psi_{ji}(x,\tau)}{\varphi_{ij}(x,\tau)}\,,
		\label{eq:ETR}
	\end{equation}
	where $I_{ij}$ is the radiation field (or specific intensity), i.e. the distribution of photons emitted or absorbed in the transition between a lower level $i$ and a higher level $j$. It depends a priori on the (reduced) frequency $x$ of the photon, the direction cosine $\mu$ and the optical depth $\tau$. In Eq.~(\ref{eq:ETR}), $A_{ji}$ and $B_{ij}$ are the Einstein coefficients associated with spontaneous emission and absorption, respectively, and $n_i$ and $n_j$ are the population densities associated with the fraction of atoms in levels $i$ and $j$, respectively. The angle-averaged emission and absorption profiles in the observer's frame $\psi_{ji}$ and $\varphi_{ij}$ are respectively defined as the convolution of the corresponding atomic emission and absorption profiles $\beta_{ji}$ and $\alpha_{ij}$ with the VDFs $f_j$ and $f_i$ of the atoms in levels $j$ and $i$:
	\begin{equation}
		\psi_{ji}(x,\tau) = \oint \frac{d\Omega}{4\pi} \int f_j(\vec{u},\tau) \beta_{ji}(\xi)d^3\vec{u} \,,
		\label{eq:psi1}
	\end{equation}
	and:
	\begin{equation}
		\varphi_{ij}(x,\tau) = \oint \frac{d\Omega}{4\pi} \int f_i(\vec{u},\tau) \alpha_{ij}(\xi)d^3\vec{u} \,,
		\label{eq:phi1}
	\end{equation}
	where $\xi$ is the frequency, in the atomic frame, of photons propagating in the direction $\vec{\Omega}$ and $d^3\vec{u}=u^2dud\vec{\Omega}_u$ is the element of integration over all atomic velocities of modulus $u$ (normalised to the most probable velocity $v_{\rm{th}}$) and direction $\vec{\Omega}_u$. For the purposes of the following discussion, the dependence on optical depth of the physical quantities is now implicit, without loss of clarity. Finally, through these profiles, a \textit{coupling} between the photon distributions $I_{ij}$ and the distributions $F_i = n_if_i$ of massive particles clearly emerges in the RTEs. Thus, the RTEs cannot be solved without simultaneously solving the set of kinetic equations for massive particles or KEEs\footnote{KEEs are also called ‘steady-state equations’ by \cite{OxeniusSimoneau94}, emphasising the fact that KEEs do not take into account the streaming of massive particles that naturally appears in Boltzmann's equations.} which, for atoms in level $i$, are written as:
	\begin{equation}
		n_i \tilde{P}_i(\vec{u}) f_i(\vec{u}) = \tilde{L}_i(\vec{u}) \,,
		\label{eq:KEE}
	\end{equation}  
	where:
	\begin{equation}
		\tilde{P}_i(\vec{u}) = Q_{V,i} + \sum_{j\neq i} \left[ A_{ij} + B_{ij}J_{ij}(\vec{u}) + C_{ij} \right] \,,
		\label{eq:defPtilde}
	\end{equation}
	is a term of ``destruction'' associated with all the processes that depopulate the atomic level $i$ in phase space, and where:
	\begin{equation}
		\tilde{L}_i(\vec{u}) = n_i Q_{V,i}f^M + \sum_{j\neq i} n_j f_j \left[  A_{ji} + B_{ji}J_{ji}(\vec{u}) + C_{ji} \right] \,,
		\label{eq:defLtilde}
	\end{equation}
	is a term of ``creation'' associated with all the processes populating the level $i$ in phase space\footnote{For notational brevity, here we use $\tilde P_i$ and $\tilde L_i$ instead of $\Pi_i$ and $L_i$ used in LPS25 (see their Eqs. (B.1)--(B.3).)}. In Eqs.~(\ref{eq:defPtilde}) and (\ref{eq:defLtilde}), we introduced the rates of inelastic collisions $C_{ij}$, velocity-changing collisions $Q_{V,i}$, and the Maxwell-Boltzmann VDF $f^M$. We also note:
	\begin{equation}
		J_{ij}(\vec{u}) = \int \alpha_{ij}(\xi)
		I_{ij} d\xi \,,
		\label{eq:Jij1}
	\end{equation}
	the partial scattering integral for the atoms in level $i$, in a velocity range between $\vec{u}$ and $\vec{u}+d\vec{u}$, and involved in the transition $i \rightarrow j$.
	 
	The integration of KEEs over all atomic velocities leads to a set of equations governing only atomic densities. We call them ``Integrated kinetic equilibrium equations'' (IKEE); they are written as:
	\begin{equation}
		n_i P_i = L_i \,,
	\end{equation}
	with:
	\begin{equation}
		P_i =  \sum_{j\neq i} \left[ A_{ij} + B_{ij}\mathcal{J}_{ij} + C_{ij} \right] \,,
	\end{equation}
	a term\footnote{Here, $P_i$ is not the same as the one used in Eq. (B.7) of \citetalias{LPS25}, as velocity-changing collisions were not taken into account.} accounting for the total amount of particle ``destruction'' in space, and:
	\begin{equation}
		L_i = \sum_{j\neq i} n_j \left[ A_{ji} + B_{ji}\mathcal{J}_{ji} + C_{ji} \right] \,,
	\end{equation}
	a term accounting for the total amount of particle ``creation'' in space. Also, $\mathcal{J}_{ij}$ is the standard scattering integral:
	\begin{equation}
		\mathcal{J}_{ij} = \int f_i(\vec{u}) J_{ij}(\vec{u}) d^3\vec{u} \,.
	\end{equation}
	
	\subsection{Emission and absorption profiles}
	
	The radiative transfer problem is indeed by nature highly coupled and is therefore difficult to solve (see e.g. \citealt{difficultRT}). The key quantities that describe the interactions between the different particle species are the emission and absorption profiles in the observer's frame. It is therefore crucial to determine these quantities self-consistently with all the kinetic equations (namely, RTEs and KEEs).   
	
	In the framework of the semi-classical picture, \citet[see also Chapters 8-10 of \citealt{HMbook}]{HOSI} phenomenologically established expressions for the atomic profiles $\alpha_{ij}$ and $\beta_{ji}$ in the context of three-level atoms, without stimulated emission. The latter can be generalised for any number of levels (see Appendix B of \citetalias{LPS25} for more details) and we have, for the atomic emission profile\footnote{A similar expression can be obtained for the atomic absorption profile $\alpha_{ij}$ by swapping indices $i$ and $j$. }:
	\begin{equation}
		\begin{split}
				& \beta_{ji}(\xi) =
				\mathrm{prob}(\rightarrow j^*) r_{ji}(\xi)
				+ \sum_{l <j}
				\mathrm{prob}(\rightarrow l^*
				\Rightarrow j)j_{lji}(\xi) \\ & +
				\sum_{k<l} \sum_{l<j} \mathrm{prob}(\rightarrow
				k^* \Rightarrow l \Rightarrow j)
				j_{klji}(\xi) \\ & + \sum_{m<k} \sum_{k<l} \sum_{l<j}
				\mathrm{prob}(\rightarrow m^*
				\Rightarrow k \Rightarrow l
				\Rightarrow j)j_{mklji}(\xi) + ... 
			\end{split}
		\label{eq:profatom}
	\end{equation}
	This expression accounts for all the processes that populate level $j$ and ultimately, lead to the transition $j \rightarrow i$. The first term involves a single photon $(\xi,\vec{\Omega})$ in the transition $j \rightarrow i$. It describes the simple emission process. The second term involves a first photon $(\vec{\Omega}',\xi')$ in the transition $l \rightarrow j$ and then a second photon $(\vec{\Omega},\xi)$ in the transition $j \rightarrow i$. When $l<j$ and $i<j$, we have Raman scattering (or resonance if $l=i$), and when $l<i<j$, we have the two-photon absorption process. The following terms, \textit{not} explicited in Eq.~(\ref{eq:profatom}), describe more complex processes involving three or more photons and, to the best of our knowledge, such mechanisms have been rarely studied in the community (see \citealt{Hubeny_3photons,Hubeny_3photons2} for more details). All these processes are described by conditional probabilities denoted by $j_{...ji}(\xi)$ (see for example Eq.~16 of \citetalias{LPS25}), carrying information about the structure of atomic levels via the generalised redistribution functions $r_{...ji}(...,\xi',\xi)$. They are weighted by their probabilities of occurrence `$\mathrm{prob}$'. The notation `$\rightarrow j^*\,$' means that the level $j$ has been naturally populated, i.e. via a process independent of the pre-existing radiation field in the medium, such as spontaneous emission, inelastic collisions or velocity-changing collisions. On the contrary, the notation `$\rightarrow l^* \Rightarrow j\,$' means that level $j$ has been non-naturally populated from a naturally populated lower level $l$, by a process dependent on the radiation field, such as absorption or stimulated emission. We recall that stimulated emission is completely neglected in this study. 
	
	Neglecting processes involving three-photon or more, and using the expressions for the probabilities `$\mathrm{prob}$' given by \cite{HOSI}, we obtain a general expression for the emission and absorption profiles in the observer's frame (see also Appendix B of \citetalias{LPS25} for more details):
	\begin{equation}
		\begin{split}
			\psi_{ji} &= 
			\frac{\varphi_{ij}}{n_j (P_j+Q_{V,j})}
			\left\{n_jQ_{V,j} \frac{\varphi_{ij}^{M*}}{\varphi_{ij}} + \sum_{p \neq j}
			n_p(A_{pj} + C_{pj})
			\frac{\varphi_{ij}^{p*}}{\varphi_{ij}}
			\right. \\ 
			& \left. + \sum_{l<j} n_lB_{lj}
			\mathcal{R}_{lji}^{l}
			 \right\} \,,
		\end{split}
		\label{eq:profGeneralExpression}
	\end{equation}
	where we have introduced the ``natural'' absorption profiles $\varphi_{ij}^*$, i.e. calculated by considering that the level $i<j$ is naturally populated\footnote{If level $i$ is naturally populated, we have $\alpha_{ij}=r_{ij}=r_{ji}$.}. We have also introduced the generalised scattering integrals $\mathcal{R}^l_{lji}$ defined as:
	\begin{equation}
		\mathcal{R}_{lji}^l(x) = \oint \frac{d\Omega'}{4\pi} \int I_{lj} \frac{R_{lji}^l(x',x)}{\varphi_{ij}(x)} dx' \,,
	\end{equation} 
	with:
	\begin{equation}
		R^l_{lji}(x',x) = \int u^2 f_l(u) \oint \frac{d\Omega}{4\pi} \oint \frac{d\Omega_u}{4\pi} r_{lji}(\xi',\xi) du \,,
	\end{equation}
	the generalised redistribution functions in the observer's frame. The superscripts $M$ and $p$ denote, respectively, that the integrations over velocities were performed using a Maxwellian and the VDF associated with atomic level $p$. Consequently, to calculate the emission profiles in the observer's frame, we must also determine, self-consistently and at every depth, the redistribution functions in the observer's frame. Then they deviate a priori from those commonly used in astrophysics (see \citealt{hummer62_redistrib} and \citealt{HubenyRedistrib}).
	
	The expression given in Eq.~(\ref{eq:profGeneralExpression}) is only valid under two assumptions: (i) $\tilde{P}_j \approx Q_{V,j} + P_j$ i.e. the number $n_jf_j\tilde{P}_j dVd^3\vec{u}$ of atoms of level $j$ that have been destroyed in the volume $dVd^3\vec{u}$ of phase space can be approximated by the fraction $f_jd^3\vec{u}$ of the total number $n_j(P_j+Q_{V,j})dV$ of atoms of level $j$ destroyed in the entire volume $dV$ ; (ii) the VDFs are assumed to be isotropic, \textit{only} for the computation of the emission profile\footnote{Anywhere else where VDFs play a role, they are angle-dependent and are, de facto, anisotropic.}, i.e. $f_j(\vec{u}) \approx f_j(u)/4\pi$ with $f_j(u)$ being the VDF integrated over all the directions of the atoms. By comparing the solutions obtained with and without assumption (i), we showed in Sect.~6 of \citetalias{LPS25} that this assumption had a minor impact on the solutions to the FNLTE radiation transfer problem. Therefore we will assume it to be reasonably valid for the case studied in this article. To quantify the isotropy of the VDFs, we introduce the parameter $a_\Omega^j$, defined for level $j$ as:
	\begin{equation}
		a_\Omega^j(\vec{u},\tau) = 4\pi \frac{f_j(\vec{u},\tau)}{f_j(u,\tau)} \,.
	\end{equation}  
	When this parameter is equal to one, the VDFs are isotropic and assumption (ii) is entirely valid. In practice, the VDFs are only slightly anisotropic. Indeed, under the conditions of \cite{SPV24}, the anisotropy factor associated with the VDF of atoms in the first excited state has, at the surface, an average value of around $1.020$ and a maximum value of around $1.20$. Under the conditions of \citetalias{LPS25}, a similar trend is observed. In the present study, deviations of the VDFs from a Maxwellian distribution are driven by the radiation field. Furthermore, when the effects of the streaming of massive particles are taken into account, the angular dependence of their distribution also becomes significant close to the surface (see \citealt{Streaming2}). Therefore, with the assumption of isotropic VDFs, we lose part of the interesting aspect of the physics particularly in the presence of global dynamical effects. Nevertheless, although the impact of these assumptions on the profiles remains to be evaluated in practice, we assume their validity for the purposes of the following discussion. To relax this assumption, more generally, one should work directly on Eqs.~(\ref{eq:psi1}) and~(\ref{eq:phi1}), namely convolve angle-dependent VDFs with the relevant atomic emission profile (also evaluated iteratively and self-consistently), thus significantly increasing the computation time. However, the more detailed angular dependence of the VDFs should be addressed in a future study.
	
	Given the expression in Eq.~(\ref{eq:profGeneralExpression}), there is a priori no reason why the emission and absorption profiles should be equal. Obtaining these profiles, and thus solving the radiation transfer problem in a physically consistent manner, can only be done by solving the set of RTE, IKEE and KEE self-consistently. In the next section, we will see how to solve numerically such a problem.       
  	
  	\section{The numerical method of solution}
  	\label{sec:MALIBU}
  	
  	When attempting to solve FNLTE radiation transfer problem, the main difficulty arises from the fact that we need to work directly on atomic velocities i.e. on a microscopic quantity. On the contrary, in the ``standard'' approach of radiative transfer, given that the VDFs are assumed to be Maxwellian, we can simply integrate over all atomic velocities, ignoring part of the physics by working with macroscopic quantities. From a numerical point of view, this change of scale has a cost, and implementing efficient iterative schemes is a challenge in itself. In the first article, we developed a naive method to solve the multi-level problem presented in Sect.~\ref{sec:Equations}. It consists of first solving the ``standard'' multi-level problem, for which all the VDFs are Maxwellian, using the MALI (Multilevel Accelerated $\Lambda$-Iteration) method from \citet[see also the code shared by \citealt{JuliaMALI}]{MALI}. This is the approximation of complete redistribution (CRD) in frequency. Then, each of the radiative quantities are updated successively in the same way as a $\Lambda$-iteration. Although the solutions obtained with such a method have been validated in the simple case of infinitely sharp three-level atoms, $\Lambda$-iteration is known for its serious convergence failures (see e.g. Chapter 12 of \citealt{HMbook}), particularly when the upper levels are naturally broadened (see e.g. \citealt{SPV24}). Thus, before moving towards greater realism, we need to develop a new numerically efficient method.    
  	
  	\subsection{MALIBU: a new approximate operator method to solve FNLTE problems}
  	
  	Approximate operator methods have been of great importance in solving ``standard'' radiative transfer problems, both in the context of the CRD approximation (ALI and MALI methods) and in more realistic cases such as partial redistribution in frequency, or PRD, with the Frequency-By-Frequency \citep[FBF,] []{FBF} and MALI-PRD methods \citep[see also $\Psi$ operators techniques of \citealt{Uitenbroek2001}]{MALI_PRD}. Because in an FNLTE radiation transfer problem, we must determine self-consistently all the VDFs associated with each of the massive particles, the emission and absorption profiles have, a priori, no reason to be identical. We are de facto in the PRD regime. 
  	Until now, only the two-level FNLTE problem has been solved using approximate operator methods. This is the Velocity-By-Velocity (UBU) method of \cite{UBU}, which is motivated by the FBF method. To solve the multi-level FNLTE problem, we generalise the MALI-PRD method to treat the KEEs for all velocities simultaneously with the RTEs and IKEEs. For this purpose, UBU method of \citet{UBU} can be appropriately generalised for the multi-level case.
  	
  	First, assuming that the populations $n_i$ and the emission and absorption profiles are known, we write the \textit{formal} solution of the RTE (see Eq.~\ref{eq:ETR}) as:
  	\begin{equation}
  		I_{ij}(x,\mu,\tau) = \Lambda_{ij}(x,\mu,\tau)[S_{ij}(x,\tau)] \,,
  		\label{eq:FormalSol}
  	\end{equation}
  	where $\Lambda_{ij}$ is an operator acting on the source function $S_{ij}$ defined as:
  	\begin{equation}
  		S_{ij}(x,\tau) = \frac{n_j A_{ji}}{n_i B_{ij}} \times \frac{\psi_{ji}(x,\tau)}{\varphi_{ij}(x,\tau)} \triangleq S_{ij}^{\rm{CRD}}(\tau)  \times \rho_{ij}(x,\tau)\,,
  		\label{eq:SourceFunction}
  	\end{equation}
  	where $S_{ij}^{\rm{CRD}} = n_jA_{ji}/n_iB_{ij}$ is a frequency-independent source function and $\rho_{ij}$ is the ratio of emission and absorption profiles.
  	The idea behind approximate operator methods is to decompose the operator as $\Lambda_{ij}^* +(\Lambda_{ij} - \Lambda_{ij}^*)$ and the source function as $S_{ij}^\dagger +(S_{ij} - S_{ij}^\dagger)$. The optimal choice for the approximate operator $\Lambda_{ij}^*$ is a diagonal operator \citep{OAB}. The formal solution given in Eq.~(\ref{eq:FormalSol}) can then be rewritten as:
  	\begin{equation}
  		I_{ij} \approx \Lambda_{ij}^* \left[S_{ij}\right] + I_{ij}^\dagger - \Lambda_{ij}^* \left[S_{ij}^\dagger\right] \,,
  		\label{eq:ApproxIntensity}
  	\end{equation}  
  	where the notation $\dagger$ refers to quantities already known from the previous iteration. To solve the transfer equation consistently with the kinetic equations of massive particles, we must propagate the decomposition of Eq.~(\ref{eq:ApproxIntensity}) in order to obtain a preconditioned KEE system, i.e. whose solution depends only on previously known quantities. The partial scattering integral defined in Eq.~(\ref{eq:Jij1}) can then be written as:
  	\begin{equation}
  		J_{ij}(\vec{u}) \approx \oint \frac{d\Omega}{4\pi}  \int \alpha_{ij}(\xi) \Lambda_{ij}^*\left[S_{ij}\right] dx + J_{ij}^{\rm{eff}}(\vec{u}) \,,
  		\label{eq:Jij_MALIBU}
  	\end{equation}     
  	with:
  	\begin{equation}
  		J_{ij}^{\rm{eff}}(\vec{u}) = J_{ij}^\dagger(\vec{u}) - \oint \frac{d\Omega}{4\pi} \int \alpha_{ij}(\xi) \Lambda_{ij}^*\left[S_{ij}^\dagger\right]dx \,, 
  		\label{eq:JijEff1}
  	\end{equation} 
  	the effective partial scattering integral, depending only on quantities that are already known. To establish a numerically efficient scheme, we now need to inject the expression of the source function into Eqs.~(\ref{eq:Jij_MALIBU}) and~(\ref{eq:JijEff1}). However, the multi-level problem that we presented in Sect.~\ref{sec:Equations} is highly non-linear. Indeed, the source function depends on emission and absorption profiles that can only be obtained by convolving an atomic profile with a VDF that is, a priori, non-Maxwellian.  Consequently, we cannot simply precondition the KEEs. To solve this problem, \cite{MALI_PRD} proposed to ``lag'' the quantity $\rho_{ij}$ by replacing it with the already known and calculated one $\rho_{ij}^\dagger$. We then have:
  	\begin{equation}
  		J_{ij}(\vec{u}) \approx \tilde{\Lambda}_{ij}^*(\vec{u})\left[S_{ij}^{\rm{CRD}}\right] + J_{ij}^{\rm{eff}}(\vec{u}) \,,
  	\end{equation}
  	with:
  	\begin{equation}
  		\tilde{\Lambda}_{ij}^*(\vec{u}) = \oint \frac{d\Omega}{4\pi} \int \alpha_{ij}(\xi) \Lambda_{ij}^* \rho_{ij}^\dagger dx \,,
  	\end{equation} 
  	an approximate diagonal operator, dependent on the atomic velocity. By inserting these last two results into Eq.~(\ref{eq:KEE}), we rewrite the left and right-hand sides of the KEE as, respectively:
  	\begin{equation}
  		n_if_i(\vec{u}) \tilde{P}_i(\vec{u}) \approx n_if_i(\vec{u})\tilde{P}_i^{\rm{eff}}(\vec{u})  + \sum_{j\neq i} \left[n_jf_iA_{ji}\tilde{\Lambda}_{ij}(\vec{u})^*\right] \,,
  		\label{eq:PreCondPi}
  	\end{equation}
  	and:
  	\begin{equation}
		\tilde{L}_i(\vec{u}) = \tilde{L}_i^{\rm{eff}}(\vec{u}) + \sum_{j\neq i} \left[n_if_j(\vec{u}) A_{ij}\tilde{\Lambda}_{ji}^*(\vec{u})	 \right]  \,,	
  	\end{equation}
  	with $\tilde{P}_i^{\rm{eff}}(\vec{u})$ and $\tilde{L}_i^{\rm{eff}}(\vec{u})$  the effective ``destruction'' and ``creation'' coefficients, i.e. calculated by replacing the partial scattering integral in Eqs.~(\ref{eq:defPtilde}) and~(\ref{eq:defLtilde}) with its effective value. For the fraction of atoms in level $i$, the KEE in Eq.~(\ref{eq:KEE}) can then be preconditioned as:
 	\begin{equation}
 		f_i(\vec{u})d_i^\dagger + \sum_{j\neq i} f_j(\vec{u}) e_{ij}^\dagger = n_i Q_{V,i} f^M \,.
 		\label{eq:KEEpreCond}
 	\end{equation}
	In the above equation, $d_i^\dagger$ is a diagonal term acting only on the VDF describing the atoms in level $i$. We identify in Eq.~(\ref{eq:PreCondPi}) that this term is simply equal to $n_i\tilde{P}_i$. The extra-diagonal terms $e_{ij}^\dagger$, acting on the VDFs of atoms in levels $j \neq i$, are given by:
	\begin{equation}
		e_{ij}^\dagger = -n_j \left[ A_{ji} + B_{ji}J_{ij}^{\rm{eff}} + C_{ji} \right] -n_iA_{ij}\tilde{\Lambda}_{ji}^*(\vec{u}) \,.
	\end{equation}  
	At this stage, the preconditioned KEEs cannot be solved because we do not know the \textit{current} population densities $n_i$ and $n_j$. We must therefore first solve the set of IKEE, which must also be preconditioned. However, the direct integration of Eq.~(\ref{eq:KEEpreCond}) over all atomic velocities also leads to non-linearities because VDFs are not known a priori. To precondition our KEEs, we assumed that $\rho_{ij} \approx \rho_{ij}^\dagger$. This means that we have tacitly assumed that the VDFs required to calculate the emission and absorption profiles were also approximated by their previously known values $f_i^\dagger(\vec{u})$. This additional assumption then allows us to fully precondition all the equations of the problem presented in Sect.~\ref{sec:Equations}, and the IKEE are written as:
	\begin{equation}
		n_iD_i^\dagger + \sum_{j\neq i} n_j E_{ij}^\dagger = 0 \,,
		\label{eq:IKEEpreCond}
	\end{equation}
	with:
	\begin{equation}
		D_i^\dagger = \sum_{j\neq i} \left[ A_{ij}\left(1-\bar{\Lambda}_{ji}^*\right) + B_{ij}\mathcal{J}_{ij}^{\rm{eff}} + C_{ij} \right]\,,
	\end{equation}
	and:
	\begin{equation}
		E_{ij}^\dagger = -\left[ A_{ji}\left(1-\bar{\Lambda}_{ij}^*\right)  + B_{ji}\mathcal{J}_{ji}^{\rm{eff}} + C_{ji} \right] \,,
	\end{equation}
	the diagonal and extra-diagonal terms of the preconditioned IKEEs, independent of atomic velocities. We have also introduced the effective ``standard'' scattering integral:
	\begin{equation}
		\mathcal{J}_{ij}^{\rm{eff}} = \int f_i^\dagger (\vec{u}) J_{ij}^{\rm{eff}} d^3\vec{u} = \mathcal{J}_{ij}^\dagger - \bar{\Lambda}_ {ij}^*\left[S_{ij}^{\dagger,\rm{CRD}} \right] \,,
	\end{equation}
	with:
	\begin{equation}
		\bar{\Lambda}_{ij}^* = \int f_i^\dagger(\vec{u}) \tilde{\Lambda}_{ij}^*(\vec{u})d^3\vec{u} \,,
	\end{equation}
	the approximated diagonal operator, independent of atomic velocities and equivalent to the one used in the MALI method. 
	
	Now, we can see that these two systems of equations are redundant. We therefore need to replace one equation to close the systems. For the IKEEs, the closing equation is the conservation of matter $\sum_i n_i = n_{\rm{tot}}$, and Eq.~(\ref{eq:IKEEpreCond}) then becomes, in matrix form, for $N$ levels atoms:
	\begin{equation}
		\begin{pmatrix}
			1 & 1 & 1 & \cdots & 1 \\
			E_{21}^\dagger & D_{2}^\dagger & E_{23}^\dagger & \cdots & E_{2N}^\dagger \\
			E_{31}^\dagger & E_{32}^\dagger & D_{3}^\dagger & \cdots & E_{3N}^\dagger \\
			\vdots & \vdots & \vdots & \ddots & \vdots \\
			E_{N1}^\dagger & E_{N2}^\dagger & E_{N3}^\dagger & \cdots & D_{N}^\dagger \\
		\end{pmatrix}
		\begin{pmatrix}
			n_1 \\
			n_2^{\vphantom{\dagger}} \\
			n_3^{\vphantom{\dagger}} \\ 
			\vdots \\
			n_N^{\vphantom{\dagger}} 
		\end{pmatrix}
		=
		\begin{pmatrix}
			n_{\rm{tot}} \\
			0^{\vphantom{\dagger}} \\
			0^{\vphantom{\dagger}} \\ 
			\vdots \\
			0^{\vphantom{\dagger}}
		\end{pmatrix} \,.
		\label{eq:IKEE_Matrix}
	\end{equation}
	For KEEs, there are two possible choices for the closure equation. Firstly, we can also define a conservation equation in phase space:
	\begin{equation}
		\sum_{i} F_i(\vec{u}) = \sum_{i} n_if_i(\vec{u}) = F_0(\vec{u}) \,,
	\end{equation}
	where $F_0$ is the distribution of all the atoms, regardless of their excitation state. Given the large number of particles present in the media we are studying, we can assume that the atoms as a whole obey a Maxwellian distribution, i.e. $F_0(\vec{u}) = n_{\rm{tot}}f^M$. However, we also know that, in the regimes considered further, we have $n_1 \gg n_i$, $\forall i\geq 2$. Thus, the velocity distribution associated with atoms in the ground state will deviate only slightly from the Maxwellian distribution, and we will have $f_1\approx f^M$. In practice, we make the latter choice, and the preconditioned KEEs defined in Eq.~(\ref{eq:KEEpreCond}) can be rewritten, in matrix form and for $N$ levels atoms as: 	
	\begin{equation}
		\begin{pmatrix}
			1 & 0 & 0 & \cdots & 0 \\
			e_{21}^\dagger & d_{2}^\dagger & e_{23}^\dagger & \cdots & e_{2N}^\dagger \\
			e_{31}^\dagger & e_{32}^\dagger & d_{3}^\dagger & \cdots & e_{3N}^\dagger \\
			\vdots & \vdots & \vdots & \ddots & \vdots \\
			e_{N1}^\dagger & e_{N2}^\dagger & e_{N3}^\dagger & \cdots & d_{N}^\dagger \\
		\end{pmatrix}
		\begin{pmatrix}
			f_1 \\
			f_2^{\vphantom{\dagger}} \\
			f_3^{\vphantom{\dagger}} \\ 
			\vdots \\
			f_N^{\vphantom{\dagger}} 
		\end{pmatrix}
		=
		\begin{pmatrix}
			1 \\
			n_2Q_{V,2}^{\vphantom{\dagger}} \\
			n_3Q_{V,3}^{\vphantom{\dagger}} \\ 
			\vdots \\
			n_NQ_{V,N} ^{\vphantom{\dagger}}
		\end{pmatrix} 
		f^M
		\,.
		\label{eq:KEE_Matrix}
	\end{equation}
	
	Finally, to solve the FNLTE radiation transfer problem, we propose the following strategy, which mainly consists of solving fully preconditioned systems of kinetic equations (IKEE and KEE) self-consistently with the kinetic equations of photons. Compared to the MALI method of \cite{MALI} or the MALI-PRD method of \cite{MALI_PRD} on which our current method is based, we solve here one MALI system for each atomic velocity. We therefore propose to name our new approximate operator method ``Multilevel Accelerated Lambda Iteration... velocity $\vec{u}$ By velocity $\vec{u}$'', that is the MALIBU method. Our iterative scheme can be summarised as follows: 
	\begin{itemize}
		\item[0:] initialise with a MALI-CRD solution (we can also initialise with LTE), which gives us the population densities, and therefore the CRD source functions. At this stage, we know $n_i^\dagger$, $f_i^\dagger$, $\varphi_{ij}^\dagger$, $\psi_{ji}^\dagger$;
		
		\item[1:] for every transition, calculate the opacity $\chi_{ij}^\dagger =n_i^\dagger B_{ij} \varphi_{ij}^\dagger $ and the emission to absorption profile ratio $\rho_{ij}^\dagger$;
		
		\item[2:] calculate the CRD source function $S_{ij}^{\dagger,\rm{CRD}}$ and the (frequency dependant) source function $S_{ij}^\dagger$;
		
		\item[3:] compute the approximated $\Lambda$ operators $\Lambda_{ij}^*(x,\mu)$, $\tilde{\Lambda}_{ij}^*$ and $\bar{\Lambda}_{ij}^*$ (which is very similar to calculating $J_{ij}$ and $\mathcal{J}_{ij}$ -- see \cite{PSP23} and \cite{SPV24} for more details);
		
		\item[4:] compute the photon distributions $I_{ij}^\dagger$ for each
		of the transitions using the previously known source function $S_{ij}^\dagger$;
		
		\item[5:] compute the partial scattering integrals
		$J_{ij}^\dagger$;
		
		\item[6:] compute the effective partial scattering integrals $J_{ij}^{\rm{eff}}$ and deduce the effective scattering integrals $\mathcal{J}_{ij}^{\rm{eff}}$ using the previoulsy known VDFs;
		
		\item[7:] update all populations $n_i$ by solving
		the preconditioned IKEE system (Eq.~{\ref{eq:IKEE_Matrix}}) at each optical depth;
		
		\item[8:] update the velocity distributions $f_i$ by
		solving the preconditioned KEE system (Eq.~{\ref{eq:KEE_Matrix}}), at
		each optical depth and atomic velocity $\vec{u}$;
		
		\item[9:] update $\varphi_{ij}$ and $\psi_{ji}$
		profiles for each of the radiatively allowed
		transitions, and return to Step 1...	
	\end{itemize}
	
	\subsection{Validation of the MALIBU method}
	
	\begin{figure}[t!]
		\includegraphics[width=\columnwidth]{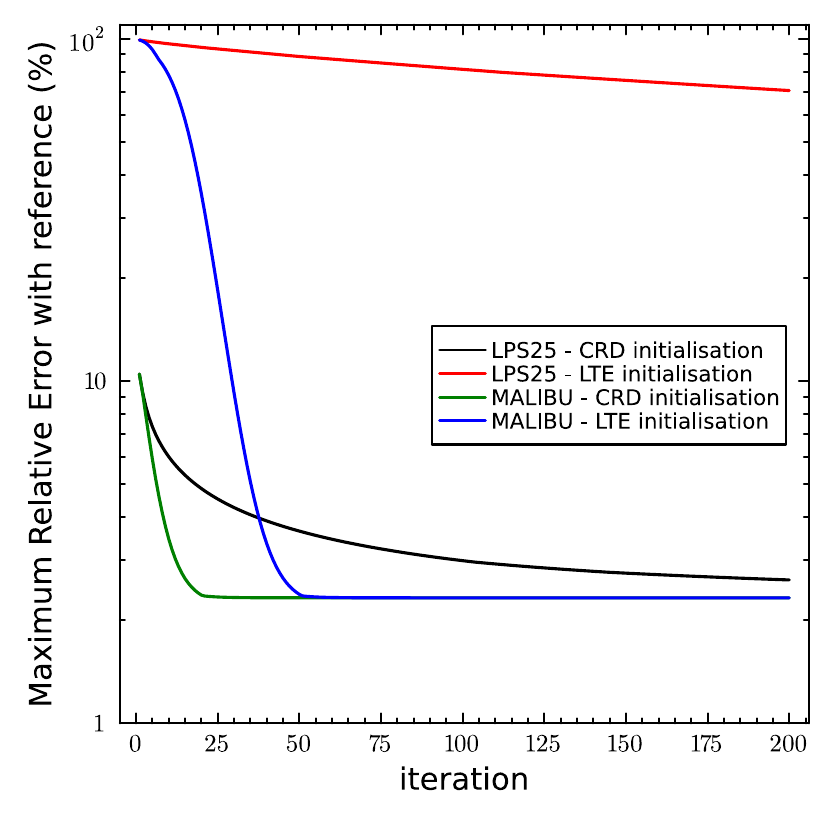}
		\caption{Maximum relative error on atomic densities (see Eq.~\ref{eq:ErrPop}), iteration after iteration with respect to the MALIBU reference solution, calculated with a high degree of precision ($8$ points per decade, $12$ cosine directions $\mu$ and $20$ azimuths). The red and black lines correspond to the solutions calculated using the \citetalias{LPS25} method, initialised at LTE and CRD respectively. The blue and green lines correspond to the solutions calculated using the MALIBU method, initialised at LTE and CRD respectively. The MALIBU method is less sensitive to the choice of initialisation and converges more rapidly than the \citetalias{LPS25} method.}
		\label{fig:Fig1}
	\end{figure}
	
	To validate our new iterative method, we must compare the results obtained with a reference solution. Apart from our own work there are, to the best of our knowledge, no other solutions to the multi-level FNLTE radiation transfer problem. Thus, as a reference, we will use the solution obtained in \citetalias{LPS25} for three-level atoms with infinitely sharp levels, without velocity-changing collisions.  
	
	We assume the same conditions as \citetalias{LPS25}, i.e. an isothermal atmosphere with temperature $T=5000\,K$, semi-infinite with 1D plane parallel geometry, sampled by a logarithmic grid with four points per decade and extending to a maximum optical depth $\tau_{\rm{max}}=10^ {14}$. The atomic model parameters are those proposed by \citet[][see also \citealt{Sampoorna2013}]{Avrett}. Angular integration is performed using six-point Gauss-Legendre quadrature for the direction cosine $\mu$ and ten-point rectangular quadrature for the azimuths. The frequency and velocity integrations are performed using trapezoidal quadrature on grids extending up to $x_{\rm{max}}=u_{\rm{max}}=4$ with a step size of $0.1$. The reference solution is then the solution obtained after $75$ iterations of the numerical scheme proposed by \citetalias{LPS25}, initialised by the CRD solution which is itself obtained using the MALI method (see again \citealt{MALI} for more details). There is no need to show these solutions again, and we refer the reader to Sect.~6 of \citetalias{LPS25} in case of need. We compare two solutions obtained by the MALIBU method with this reference. One is initialised using the CRD solution. The other is initialised at the LTE. We will also use another solution obtained with the method of \citetalias{LPS25} and initialised at LTE.  
	
	To compare these solutions with the reference, we introduce a relative error term between the population densities $n_i$, calculated in one of the three situations described above, and the populations $n_i^{\rm{ref}}$ associated with the reference:
	\begin{equation}
		{\rm{Err}}_{\rm{rel},i} = \left\lvert \frac{n_i - n_i^{\rm{ref}}}{n_i^{\rm{ref}}} \right\lvert \,.
		\label{eq:ErrPop}
	\end{equation}
	Firstly, the iterative method of \citetalias{LPS25} clearly exhibits the pathological characteristics of $\Lambda$-iteration. Indeed, when initialised at LTE, the relative error defined in Eq.~(\ref{eq:ErrPop}) has an average of about $40\%$ after $75$ iterations. Further iterations do not improve the results. This method does not converge for any initial conditions, or converges very slowly. There is no guarantee that this method, even when initialised at CRD, will converge towards the solution in more physically realistic situations. On the other hand, regardless of the choice of the initialisation, the MALIBU method reproduces the reference solution with very good accuracy, with an average and maximum relative error on populations of approximately $0.071\%$ and $0.41\%$. This method is therefore a priori much more robust.
	
	Now, we consider the reference solution to be the solution obtained with MALIBU, initialised at the CRD. This solution is computed for a very fine set of parameters, i.e. 8 points per decade and a finer angular quadrature with 12 directions cosine $\mu$ and 20 azimuths. Only $300$ iterations are necessary to achieve convergence. Beyond that, the updated atomic densities no longer change in comparison with the previous iteration. In Fig.~\ref{fig:Fig1}, we show iteration after iteration the maximum relative error with respect to this new reference, for the MALIBU method and for the method previously used in \citetalias {LPS25}. We observe that the MALIBU method converges in about 50 iterations when initialised at LTE (blue line) and in only 20 iterations when initialised at CRD (green line). On the contrary, the method of \citetalias{LPS25} does not converge, as we specified above, when we initialise at LTE (red line) and converges slowly when we initialise at CRD (black line).  
	
	Unlike the method described in \citetalias{LPS25}, we must compute an approximate operator $\tilde{\Lambda}_{ij}^*(\vec{u})$ for each atomic line and at each iteration. Consequently, a MALIBU iteration is more computationally expensive than an iteration of the method described in \citetalias{LPS25}. Our MALIBU method, fully implemented in \textit{Julia}, takes $\approx 0.8\,s$ to complete a single iteration, whereas the method of \citetalias{LPS25} takes $\approx 0.5\,s$.
	However, our new method has two major advantages over the method described in \citetalias{LPS25}. It converges independently of the choice of the initialisation and, under the same conditions, converges in significantly fewer iterations than the previous method.
	
	In what follows, the solutions presented are calculated using this new MALIBU method, which we have just validated.     
	
	\section{An atomic model with natural broadening}
	\label{sec:CaII_model} 
	
	\tikzset{every picture/.style={line width=0.75pt}} 
	
	\subsection{Broadening of spectral lines in the framework of the Weisskopf-Woolley model}
	
	\begin{table}[t!]               
		\centering  
		   \caption{Ca\,{\sc ii} atomic parameters}                    
		\begin{tabular}{c c c c c c}      
			\hline\hline               
			$j$ & $i$ & $A_{ji}$ & $C_{ji}$ & $\nu_{0,ji}$ & $a_{ji}$ \\   
			& & (s$^{-1}$) & (s$^{-1}$)& Hz & \\      
			\hline                     
			$2$ & $1$ & Forbidden &               $1.31 \times 10^{4}$ & $4.10\times10^{14}$ & ... \\   
			$3$ & $1$ & Forbidden &              $1.31 \times 10^{4}$ & $4.12\times10^{14}$ & ... \\ 
			$4$ & $1$ & $1.40\times10^8$ &   $2.23 \times 10^{4}$ & $7.56\times10^{14}$ & $3.31\times10^{-3}$ \\ 
			$5$ & $1$ & $1.40\times10^8$ &   $2.23 \times 10^{4}$ & $7.63\times10^{14}$ & $3.28\times10^{-3}$ \\ 
			$3$ & $2$ & Forbidden & 	         $3.27 \times 10^{4}$ & $2.00\times10^{12}$ & ... \\ 
			$4$ & $2$ & $1.06\times10^7$ & $1.00 \times 10^{5}$ & $3.46\times10^{14}$ & $7.23\times10^{-3}$ \\ 
			$5$ & $2$ & $1.11\times10^6$ &   $2.24 \times 10^{4}$ & $3.53\times10^{14}$ & $7.09\times10^{-3}$ \\ 
			$4$ & $3$ & Forbidden & 			  $1.59 \times 10^{4}$ & $3.44\times10^{14}$ & ... \\ 
			$5$ & $3$ & $9.60\times10^6$ &   $9.88 \times 10^{4}$ & $3.51\times10^{14}$ & $7.13\times10^{-3}$\\ 
			$5$ & $4$ & Forbidden & 			  $2.33 \times 10^{4}$ & $7.00\times10^{12}$ & ... \\ 
			\hline                             
		\end{tabular}
		\tablefoot{Atomic parameters for the Five-level Ca\,{\sc ii} atomic model with H\,\&\,K lines and infrared triplet. Statistical weights for the first five levels of Ca\,{\sc ii} are: $g_1 = 2$, $g_2 = 4$,
			$g_3 = 6$, $g_4 = 2$, and $g_5 = 4$. Table taken from \cite{Sampoorna2013}. The damping parameters have been recomputed. }   
		\label{table:1} 
	\end{table}
	
	Consider an atom in excitation level $i$, with energy $E_i$ relative to the ground state, or frequency $\xi_i=E_i/h$ in the atomic frame. Numerous interactions, radiative or not, may excite or de-excite this atom to a level $j$ with energy $E_j \neq E_i$. Then, on average, this atom will remain in level $i$ during a characteristic lifetime. According to Heisenberg's uncertainty principle on time-energy, this results in a small variation of the energy of the atomic level itself. This is the lifetime or natural broadening of atomic levels. Another source of spectral line broadening is collisions between a radiating atom, behaving like an oscillator, and a perturber (see \citealt{HMbook}, Chapters 8-10 for more details). These are referred to as phase-changing collisions. They are responsible for what is known as pressure or collisional broadening.   
	
	We propose to include these line broadening mechanisms for the first time within the framework of the multi-level FNLTE formalism presented above. In the semi-classical approach or Weisskopf-Woolley model (\citealt{weisskopf1931,weisskopf1933,woolley1931,woolley1938} -- see also e.g. \citealt[Chapters 8-10,]{HMbook} or \citealt{OxeniusSimoneau94}), the broadening of atomic levels is described as a continuum of sublevels. This notion applies only to a \textit{set} of atoms, and these sublevels should not be confused with ``true'' atomic sublevels associated. For a set of atoms of the same type, the number of sublevels between $\xi_i$ and $\xi_i+d\xi_i$ is given by $G_i(\xi_i)d\xi_i$, where $G_i$ is a normalised weight function. Typically, $G_i$ is a Lorentzian:
	\begin{equation}
		\mathcal{L}_i(\xi_i-\xi_i^c) = \left(\frac{\delta_i}{\pi}\right) \frac{1}{\delta_i^2 + (\xi_i - \xi_i^c)^2} \,,
		\label{eq:Lorentzian}
	\end{equation}      
	centred on the most probable frequency $\xi_i^c$ with a damping parameter $\delta_i$ homogeneous to the lifetime of level $i$. This parameter accounts for all the processes able to depopulate level $i$ or reorganise its internal sublevel structure. Spontaneous emission, radiative absorption and inelastic collisions depopulate level $i$ by inducing a transition to a level $j \neq i$. Velocity-changing collisions (v.c.c) also depopulate level $i$ in phase space. Phase-changing collisions (p.c.c) induce transitions between sublevels of a single level $i$. The rates of these last two collisions are all or part of the total elastic collision rate $Q_E=Q_V+Q_P$, where $Q_V$ is the v.c.c rate and $Q_P$ is the p.c.c rate. Finally, the damping parameter is written as:
	\begin{equation}
		4\pi \delta_i = Q_E + \sum_{l <i} A_{il} + \sum_{u>i} B_{iu}J_{iu} + \sum_{j\neq i} C_{ij} \,.
		\label{eq:damping_parameter}
	\end{equation}
	In practice, inelastic collisions and radiative absorption are negligible compared to spontaneous emission, and therefore we can write:
	\begin{equation}
		4\pi \delta_i \approx Q_E + \sum_{l <i} A_{il} \,.
		\label{eq:damping_parameter2}
	\end{equation}
	While the distribution of sublevels $G_i$ is known in advance, the way in which a set of atoms occupies these sublevels is, a priori, unknown. Without going into details, which were extensively covered by \cite{OxeniusSimoneau94}, determining this occupation state requires solving a kinetic equation directly at the sublevel scale. Without explicitly solving these equations, the expressions for the atomic profiles obtained phenomenologically by \cite{HOSI} and recalled in Eq.~(\ref{eq:profatom}) allow us to treat this broadening of the atomic levels in an equivalent way. However, this requires defining ab initio a set of generalised redistribution functions that we will not define here in general terms (for more details, see e.g. \citealt{hummer62_redistrib,HubenyRedistrib}).     
	
	\subsection{H\,\&\,K lines and infrared triplet of Ca\,{\sc ii}}
	
	\begin{figure}[t]
		\begin{center}
			\fbox{\begin{tikzpicture}[x=0.75pt,y=0.75pt,yscale=-0.8,xscale=0.9]
					\draw [color={rgb, 255:red, 140; green, 14; blue, 238 }  ,draw opacity=1 ]   (207.46,120.82) -- (207.66,335.27) ; 
					\draw [color={rgb, 255:red, 140; green, 14; blue, 238 }  ,draw opacity=1 ]   (308.11,81.48) -- (207.66,335.27) ;
					\draw [color={rgb, 255:red, 238; green, 14; blue, 14 }  ,draw opacity=1 ]   (308.11,81.48) -- (308.01,261.86) ;
					\draw [color={rgb, 255:red, 238; green, 14; blue, 14 }  ,draw opacity=1 ]   (207.46,120.82) -- (308.01,261.86) ;
					\draw [color={rgb, 255:red, 238; green, 14; blue, 14 }  ,draw opacity=1 ]   (308.11,81.48) -- (412.66,207.07) ; 
					\draw [line width=1.5]    (173.28,335.29) -- (242.04,335.25) ;
					\draw [line width=1.5]    (267.31,261.91) -- (336.07,261.87) ; 
					\draw [line width=1.5]    (366.54,207.63) -- (435.3,207.59) ;] 
					\draw [line width=1.5]    (174.4,121.2) -- (240.61,121.2) ;
					\draw [color={rgb, 255:red, 128; green, 128; blue, 128 }  ,draw opacity=1 ]   (174.4,123.21) -- (240.61,123.21) ;
					\draw [color={rgb, 255:red, 128; green, 128; blue, 128 }  ,draw opacity=1 ]   (174.4,124.83) -- (240.61,124.83) ; 
					\draw [color={rgb, 255:red, 128; green, 128; blue, 128 }  ,draw opacity=1 ]   (174.4,126.84) -- (240.61,126.84) ;
					\draw [color={rgb, 255:red, 128; green, 128; blue, 128 }  ,draw opacity=1 ][fill={rgb, 255:red, 155; green, 155; blue, 155 }  ,fill opacity=1 ]   (174.4,129.85) -- (240.61,129.85) ; 
					\draw [color={rgb, 255:red, 128; green, 128; blue, 128 }  ,draw opacity=1 ]   (240.62,118.91) -- (174.42,119.4) ;
					\draw [color={rgb, 255:red, 128; green, 128; blue, 128 }  ,draw opacity=1 ]   (240.62,117.29) -- (174.41,117.78) ;
					\draw [color={rgb, 255:red, 128; green, 128; blue, 128 }  ,draw opacity=1 ]   (240.62,115.28) -- (174.41,115.77) ;
					\draw [color={rgb, 255:red, 128; green, 128; blue, 128 }  ,draw opacity=1 ]   (240.61,112.26) -- (174.4,112.75) ;
					\draw [line width=1.5]    (270.16,82.09) -- (336.37,82.09) ;
					\draw [color={rgb, 255:red, 128; green, 128; blue, 128 }  ,draw opacity=1 ]   (270.16,84.1) -- (336.37,84.1) ;
					\draw [color={rgb, 255:red, 128; green, 128; blue, 128 }  ,draw opacity=1 ]   (270.16,85.72) -- (336.37,85.72) ;
					\draw [color={rgb, 255:red, 128; green, 128; blue, 128 }  ,draw opacity=1 ]   (270.16,87.73) -- (336.37,87.73) ; 
					\draw [color={rgb, 255:red, 128; green, 128; blue, 128 }  ,draw opacity=1 ][fill={rgb, 255:red, 155; green, 155; blue, 155 }  ,fill opacity=1 ]   (270.16,90.75) -- (336.37,90.75) ;
					\draw [color={rgb, 255:red, 128; green, 128; blue, 128 }  ,draw opacity=1 ]   (336.38,79.8) -- (270.17,80.29) ;
					\draw [color={rgb, 255:red, 128; green, 128; blue, 128 }  ,draw opacity=1 ]   (336.38,78.18) -- (270.17,78.67) ; 
					\draw [color={rgb, 255:red, 128; green, 128; blue, 128 }  ,draw opacity=1 ]   (336.38,76.17) -- (270.17,76.66) ;
					\draw [color={rgb, 255:red, 128; green, 128; blue, 128 }  ,draw opacity=1 ]   (336.37,73.15) -- (270.16,73.65) ;
					
					\draw (183.79,266.31) node [anchor=north west][inner sep=0.75pt]  [rotate=-270.26]  {$3963\AA \ ( H)$};
					\draw (206.79,273.18) node [anchor=north west][inner sep=0.75pt]  [rotate=-291.63]  {$3934\AA \ ( K)$};
					\draw (278.12,185.09) node [anchor=north west][inner sep=0.75pt]  [rotate=-53.67]  {$8662\AA $};
					\draw (328.94,138.48) node [anchor=north west][inner sep=0.75pt]  [rotate=-89.28]  {$8498\AA $};
					\draw (367.46,117.8) node [anchor=north west][inner sep=0.75pt]  [rotate=-52.01,xslant=0.05]  {$8542\AA $};
					\draw (147.8,321.89) node [anchor=north west][inner sep=0.75pt]    {$( 1)$};
					\draw (148.9,106.24) node [anchor=north west][inner sep=0.75pt]    {$( 4)$};
					\draw (244.93,70.55) node [anchor=north west][inner sep=0.75pt]    {$( 5)$};
					\draw (431.59,193.72) node [anchor=north west][inner sep=0.75pt]    {$( 3)$};
					\draw (332.73,247) node [anchor=north west][inner sep=0.75pt]    {$( 2)$};

			\end{tikzpicture}}
			\caption{Schematic view of the atomic model associated with the H\,\&\,K lines (purple lines) and the infrared triplet (red lines) of Ca\,{\sc ii}. The atomic levels are numbered in increasing order of energy. This figure is a slightly modified version of the one shown in Fig. 1 of \cite{Sampoorna2013}.}
			\label{fig:Fig2}
		\end{center}
	\end{figure}
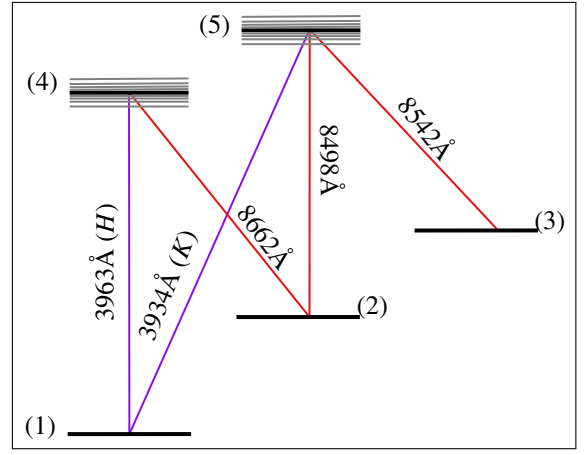
	
	We now propose to deal with the H\,\&\,K lines and the infrared triplet of Ca\,{\sc ii}. In addition to its wide use in astrophysics, the choice of this atomic model as the first study of a realistic multi-level FNLTE problem is motivated by its relative simplicity. Indeed, in this five-level model (see Fig.~\ref{fig:Fig2}) the levels $2$ and $3$ are metastable and behave like ground states. Only the two upper levels are broadened, and only five transitions are \textit{radiatively} allowed. The atomic data used are compiled in Table~\ref{table:1} based on the models of \citet[see also \citealt{Sampoorna2013}]{Avrett}. The transition frequencies $\nu_{0,ij}=\lvert \xi_j^c - \xi_i^c \lvert$ and the values of the Einstein coefficients for spontaneous emission are those of the National Institute of Standards and Technology (\citetalias{NIST_ASD}) database. The lines denoted as ``forbidden'' have very low Einstein coefficients (of the order of unity) compared to those of radiatively allowed lines. For a temperature of $5000\,K$ and an electron density of $7.664\times10^{10}$cm$^{-3}$, the inelastic collision rates have been estimated by \cite {shine_linsky_1974}. For each of the transitions $i \leftrightarrow j$ with $j>i$ and Doppler width $\Delta_{ij}$, a dimensionless damping parameter $a_{ji} = (\delta_j + \delta_i)/\Delta_{ij}$ is defined. In practice, the ground state and metastable states are infinitely sharp, i.e. $\delta_i=0$, so that $a_{ji}=a_j=\delta_j/\Delta_{ij}$ where $\delta_{j}$ is defined in Eq.~(\ref{eq:damping_parameter2}). The values of these parameters given in Table~\ref{table:1} were calculated without elastic collisions. They differ from those provided by \cite{Sampoorna2013}, who considered that the damping parameter was only given by $\delta_j = A_{ji}$. 
	
	\subsection{Redistribution in Calcium} 
	
	In the five-level model of Ca\,{\sc ii} described above, we study the formation of five spectral lines describing the transitions $1 \leftrightarrow 5$, $2 \leftrightarrow 5$, $3 \leftrightarrow 5$, $1 \leftrightarrow 4$ and $2 \leftrightarrow 4$. For each of these transitions, the lower level is naturally populated because there is no radiative absorption process populating these levels from a lower level. Therefore, neither two-photon absorption nor three-photon or more processes exist. The atomic absorption profiles $\alpha_{ij}$ are therefore all equal to the generalised one-photon redistribution function $r_{ij}$, i.e. a Lorentzian (see \citealt{OxeniusSimoneau94}):
	\begin{equation}
		r_{ij}(\xi) = \left(\frac{\delta_{j}}{\pi}\right) \frac{1}{\delta_{j}^2 + (\xi - \nu_{0,ij})^2} \,.
	\end{equation} 
	In the observer's frame, the absorption profiles are given by Eq.~(\ref{eq:phi1}) and we have:
	\begin{equation}
		\varphi_{ij}(x) = \int_0^\infty \frac{uf_i(u)}{2\pi} \left[\arctan \left(\frac{u+x}{a_{ji}}\right) + \arctan \left(\frac{u-x}{a_{ji}}\right) \right] du \,.
		\label{eq:phifl}
	\end{equation}
	When $f_i=f^M$, this expression of the absorption profile is equivalent to a Voigt profile\footnote{Note that $\varphi_{ij} = \varphi_{ij}^{i*}$.}. Also, since the atomic absorption profile is a Lorentzian, evaluating the partial scattering integrals defined in Eq.~(\ref{eq:Jij1}) is more tedious than in \citetalias{LPS25}. In practice, we use the numerically costly ``Pole-Away'' method from \citet[see also \citealt{SPV24}]{Bommier1997a,Bommier1997b}.   
	
	Since there is no three-photon process, the emission profiles in the observer's frame are given by Eq.~(\ref{eq:profGeneralExpression}). However, to calculate them, we need to know as many generalised redistribution functions as there are radiatively allowed transition sequences $l \rightarrow j \rightarrow i$ (with $l<j$). In total, thirteen of these functions must be determined, at each iteration, in a self-consistent way. In the atomic frame, there are two distinct redistribution processes. If, for a set of atoms, there is an elastic collision that can induce transitions between the sublevels of a level $j$ between the time when it is populated and the time when it is depopulated, then the redistribution process is said to be decorrelated. The frequency of the photon emitted in the transition $l \rightarrow j$ does not determine the frequency of the one emitted in the transition $j \rightarrow i$. On the contrary, if such collisions do not occur, the redistribution process is said to be correlated. The generalised redistribution functions are then, in the atomic frame, a linear combination of a purely correlated redistribution function $r^{\rm{II}}$ and a purely decorrelated one $r^{\rm{III}}$ \citep{OmontSmithCooper}\footnote{This expression is valid for the Ca\,{\sc ii} case studied here but is not a general expression applicable to all atoms.}:
	\begin{equation}
		r_{lji}(\xi',\xi) = \gamma_j r_{lji}^{\rm{II}}(\xi',\xi) + (1-\gamma_j)r_{lji}^{\rm{III}}(\xi',\xi) \,,
	\end{equation}  
	where the branching ratio $\gamma_j$ is given by (see e.g. \citealt{HMbook} or \citealp{hubenycooper1986}):
	\begin{equation}
		\gamma_j = \frac{P_j + Q_V}{P_j + Q_E} \,.
		\label{eq:gamma}
	\end{equation} 
	\citet[see also \citealt{hummer62_redistrib}]{HubenyRedistrib} gives an expression for the generalised redistribution functions $r^{\rm{III}}$ and $r^{\rm{II}}$:
	\begin{equation}
		r_{lji}^{\rm{III}}(\xi',\xi) = \mathcal{L}_j(\xi'-\nu_{0,lj}) \times \mathcal{L}_j(\xi-\nu_{0,ji}) \,,
	\end{equation}
	and
	\begin{equation}
		r_{lji}^{\rm{II}}(\xi',\xi) = \mathcal{L}_j(\xi'-\nu_{0,lj}) \times \delta \left([\xi'-\nu_{0,lj}] - [\xi-\nu_{0,ji}]\right)\,.
	\end{equation}
	In the observer's frame, it can be shown that the generalised redistribution function is also given by a linear combination of two distinct redistribution mechanism:
	\begin{equation}
		R^l_{lji}(x',x) = \gamma_j R_{lji}^{l,\rm{II}}(x',x) + (1-\gamma_j) R_{lji}^{l,\rm{III}}(x',x) \,,
		\label{eq:R}
	\end{equation}
	with:
	\begin{equation}
		\begin{split}
			R_{lji}^{l,\rm{III}}(x',x) = \int_0^\infty &\frac{f_l(u)}{4\pi^2} 
			\left[
			\arctan\left(\frac{x+u}{a_j}\right) - \arctan\left(\frac{x-u}{a_j}\right)
			\right]
			\\
			&\times \left[
			\arctan\left(\frac{x'+u}{a_j'}\right) - \arctan\left(\frac{x'-u}{a_j'}\right)
			\right]
			du \,,
		\end{split}
		\label{eq:RIII}
	\end{equation}
	where $a_j=\delta_j/\Delta_{ji}$ and $a_j'=\delta_j/\Delta_{lj}$. The generalised redistribution function $R^{l,II}$ is written as (see also \citealt{Uitenbroek}, who derived this expression for $f_l=f^M$):
	\begin{equation}
		R_{lji}^{l,\rm{II}}(x',x) = \int_{u_{\min}}^\infty \frac{f_l(u)}{4\pi} 
		\left[
		\arctan\left(y_{\max}\right) - \arctan\left(y_{\min} \right)
		\right] du \,,
		\label{eq:RII}
	\end{equation}
	with:
	\begin{equation}
		u_{\min} = \frac{\lvert x - \alpha x' \lvert}{1+\alpha} \,,
	\end{equation}
	where $\alpha = \Delta_{lj}/\Delta_{ji}$ and:
	\begin{equation}
		y_{\max} = \min\left(\frac{(u-x)}{a_j} ; \alpha \frac{(u-x')}{a_j}\right) \,,
	\end{equation}
	and
	\begin{equation}
		y_{\min} = \max\left(\frac{-(u+x)}{a_j} ; -\alpha \frac{(u+x')}{a_j}\right) \,.
	\end{equation}
	A method of integration is proposed in Appendix~\ref{app:AppA} to determine, numerically, the absorption profiles and these generalised redistribution functions (Eqs.~\ref{eq:phifl},~\ref{eq:RIII} and~\ref{eq:RII}).
	
	\section{Full non-LTE treatment of H\,\&\,K lines and infrared triplet of Ca\,{\sc ii}}
	\label{sec:Results}
	
	In this section, we present solutions to the multi-level FNLTE radiation transfer problem applied to the five-level Ca\,{\sc ii} model described above. We consider a 1D plane parallel atmosphere with a maximum optical depth of $\tau_{\rm{max}}=10^{14}$ for the $1 \leftrightarrow 4$ transition (H line). The optical depth grid is sampled on a logarithmic scale, with $4$ points per decade and an initial step size of $10^{-3}$. Integrations over the direction cosine $\mu$ are performed using a six-node Gauss-Legendre quadrature. The integrations over the azimuths, necessary to evaluate the scattering integral (see \citealp{PSP23,SPV24} for more details), are performed with a simple rectangular quadrature with $10$ nodes. The normalised velocity grid extends from $0$ to $u_{\rm{max}}=6$ with a step size of $0.1$. The frequency grid must correctly resolve both the core and the wings of the line under study. We choose $x_{\rm{max}}=1000$ such that, for any transition $i \leftrightarrow j$, we have $\varphi(x_{\rm{max}}) \approx \varphi_{D}(x_{D,\rm{max}}=4)$, with $x_{D,\rm{max}}$ being the choice of maximum frequency when the absorption profile is a Doppler profile $\varphi_D(x)=\pi^{-1/2}e^{-x^2}$. Then, we use a composite frequency grid, regular in the core from $0$ to $x_{D,\rm{max}}=4$ with a step of $0.1$ and then logarithmic up to $x_{\rm{max}}$ with $50$ points. The frequency and velocity integrations are performed using a trapezoidal rule. The FNLTE solutions are obtained using our new MALIBU method presented in Sect.~\ref{sec:MALIBU}. The iterative process is initialised with the multi-level CRD solution, obtained after $100$ iterations of the MALI-CRD method (see again the code shared by \citealt{JuliaMALI}).           
	
	\subsection{Benchmarking against Cross-Redistribution (XRD)}
	
 	Before presenting our results for the FNLTE treatment of Ca\,{\sc ii}, we must validate our tools and methods. Here, we demonstrate that we are able to reproduce the results of the so-called ``Cross-Redistribution'' (XRD). In XRD, all velocity distributions required for calculating absorption profiles (Eq.~\ref{eq:phifl}) and redistribution functions (Eqs.~\ref{eq:RIII} and~\ref{eq:RII}) are a priori Maxwellians. By applying this hypothesis and consequently partially degrading our FNLTE formalism, we should obtain the same results as for XRD.
 	
 	The XRD solution used as a reference is obtained after $200$ iterations of the MALI-XRD method from \cite{Sampoorna2013}, directly based on the MALI-PRD method from \cite{MALI_PRD}. On the other hand, by applying the XRD assumptions, we calculate the solution obtained after $200$ iterations of our MALIBU method. The results are reproduced with satisfactory accuracy. Indeed, the relative error defined in Eq.~(\ref{eq:ErrPop}) between the population densities calculated by the MALI-XRD method (reference) and our MALIBU method is only of $0.24\%$ on average and of $1.77\%$ at maximum. Such an error can also be defined on the source functions for which we observe a maximum relative error of $6.20\%$. This more important error can be explained by the radically different ways of evaluating some radiative quantities in FNLTE, in particular the scattering integrals $\mathcal{J}_{ij}$. Indeed, these are calculated in the ``standard'' formalism by integrating the radiation field over all frequencies, with a priori known absorption profile. In FNLTE, this same quantity is obtained by integrating, on velocities, the partial scattering integral $J_{ij}(\vec{u})$ which is itself evaluated numerically, introducing potential additional errors. On the other hand, this error on the source functions is relatively small on average, with a value of $0.81\%$. We can therefore consider that our tools, developed specifically for the FNLTE formalism, are sufficiently reliable to be applied to a real, non-degraded problem. However, we must keep in mind that our integration methods will need to be improved in our future work in order to further reduce these discrepancies.
	
	\subsection{FNLTE calculations without elastic collisions}
	
	\begin{figure}[t!]
		\includegraphics[width=\columnwidth]{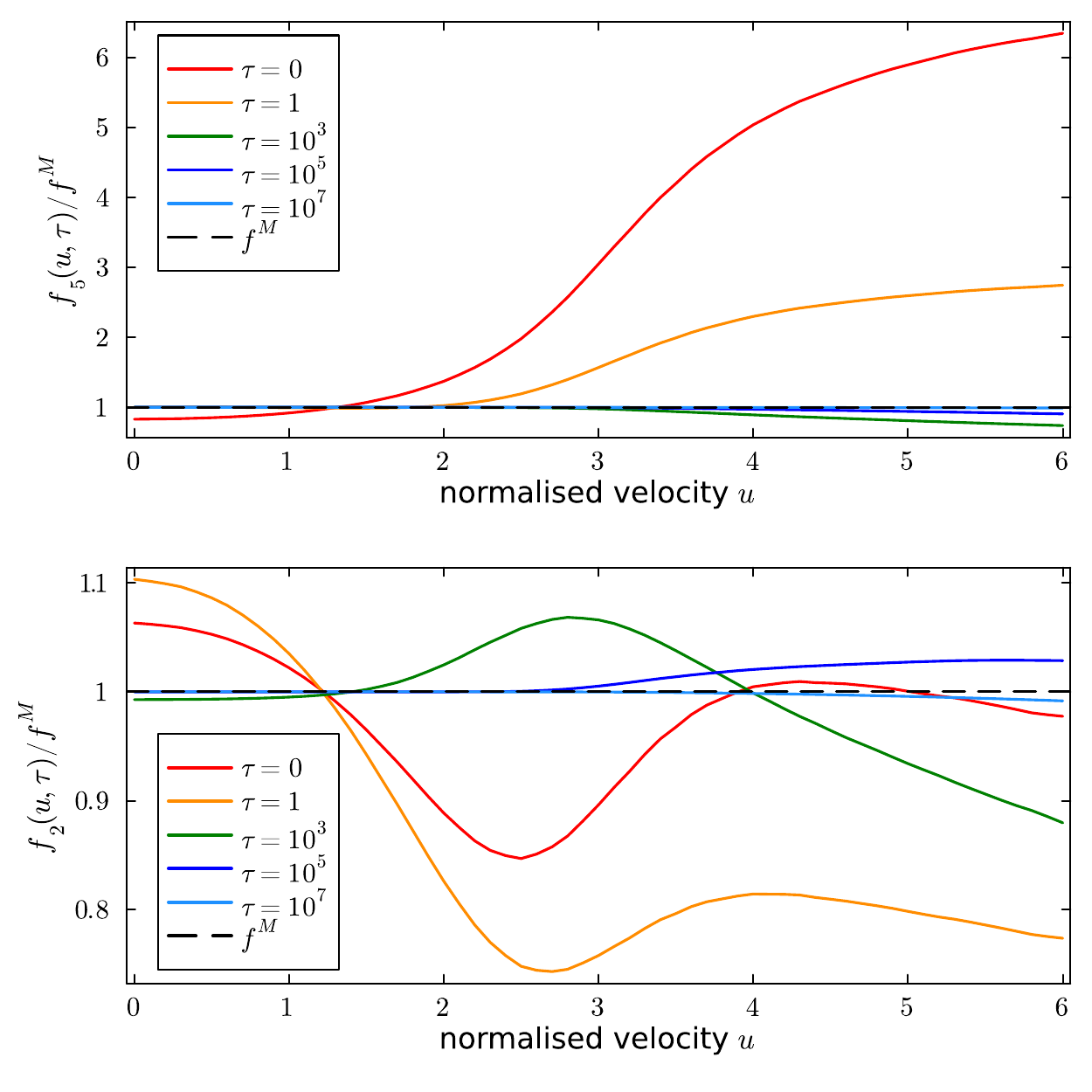}
		\caption{Velocity distributions functions, normalised to the Maxwellian, associated with the fraction of atoms in levels $5$ (top) and $2$ (bottom). Their variation is shown for different optical depths $\tau=0,1,10^3,10^5,10^7$. The Maxwellian distribution is shown as a dashed line. Clearly, the VDFs calculated in FNLTE are not Maxwellians.}
		\label{fig:Fig3}
	\end{figure}
	
	For the same atmospheric model, we now study the solutions to the FNLTE problem, applied to Ca\,{\sc ii}. Firstly, we do not take elastic collisions into account. Unlike in the previous case, we now determine all VDFs self-consistently. Only our approach is able to provide this new quantities. Fig.~\ref{fig:Fig3} shows the VDFs of the excited atoms in level $5$ (top) and level $2$ (bottom). Similar results are obtained for $f_4$ and $f_3$. Clearly, the velocity distributions are not Maxwellians, and deviations from Maxwellianity increase as we approach the surface. This result is expected because, deep in the interior, the atmosphere is completely thermalised, i.e. $f_i \rightarrow f^M$ when $\tau \gg 1$. 
 
	However, we note that deviations from the Maxwellian of the VDFs associated with metastable levels are smaller than those associated with upper levels. To provide an explanation, we must note that it is the interaction between the radiation field and matter that leads to these deviations. Indeed, in the observer's frame, only atoms with a particular velocity will be able to absorb a photon of a given frequency, due to the Fizeau-Doppler effect. The radiation field therefore ``selects'' a particular population of atoms, all of which have a common dynamic that is correlated with the properties of the absorbed photons. Strictly speaking, this new population of excited atoms cannot obey the same distribution as the initial one. This is the mechanism involved for upper levels, as these are non-naturally populated from lower levels via radiative absorption. In addition, we know that everything that populates this higher level depopulates the corresponding lower level. Thus, if a population with a common dynamics has been \textit{removed} from the initial population of atoms in the metastable levels, we would expect the distribution of these levels to also be perturbed compared to the initial one\footnote{This is exactly what is described by the conservation equation $\sum_i F_i = F_0$.}. However, this effect is not significant in the case studied here because spontaneous emission from the upper level tends to establish a natural population for the metastable levels which are also, approximately $100$ times more numerous than those in the upper levels. Therefore, in practice, VDFs of the metastable levels deviates only slightly from the Maxwellian distribution, as we observe in bottom panel of Fig.~\ref{fig:Fig3}. As expected, metastable levels behave like additional ground states.  
	
\begin{figure}[t!]
		\includegraphics[width=\columnwidth]{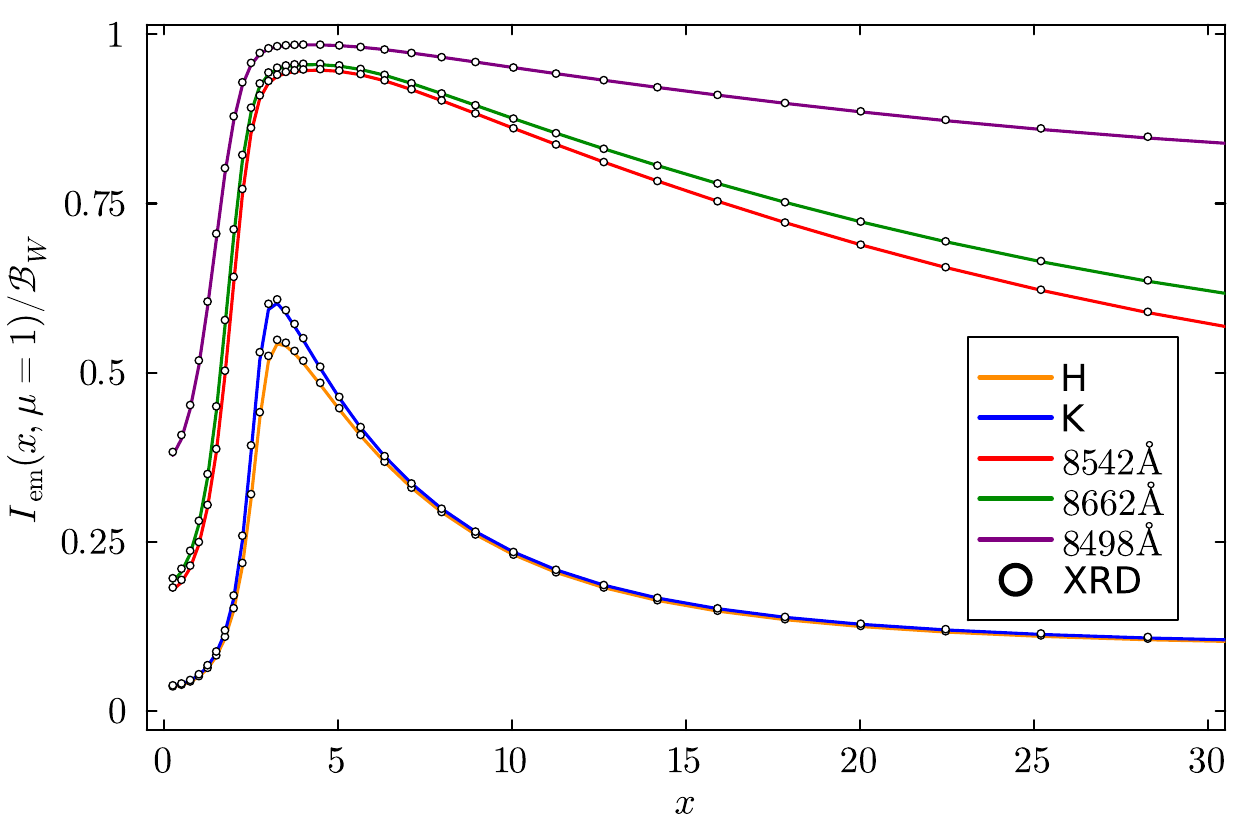}
		\caption{Frequency variation of the emergent intensity at $\mu=1$, normalised to the Wien function $\mathcal{B}_W$, for the H\,\&\,K lines and the infrared triplet of Ca\,{\sc ii}. The coloured lines are the results obtained with the FNLTE formalism, using the MALIBU method. These results are compared with those of the XRD (circles).}
		\label{fig:Fig4}
	\end{figure}

	We recall that the XRD assumptions consist precisely in assuming that the VDFs associated with the lower levels, used to calculate the absorption profiles and generalised redistribution functions, are Maxwellian. Consequently, we can expect the solutions to the FNLTE problem to deviate only slightly from the XRD solutions. Fig.~\ref{fig:Fig4} shows the emergent intensities associated with the H\,\&\,K lines and the infrared triplet of Ca\,{\sc ii}, calculated using FNLTE (coloured lines) and XRD (circles). The average relative difference between these two solutions is only $1.30\%$ for the H\,\&\,K lines, and of the order of one percent for the infrared triplet. Clearly, under the physical conditions of the atmosphere studied, we do not find evidence of pure FNLTE effects on the formation of Ca\,{\sc ii} spectral lines. Far from providing a general conclusion, we must nevertheless, in the future, inspect the physical conditions and atomic models leading to significant deviations in the VDFs of the lower levels and, inevitably, leading to deviations in the solutions from those of the ``standard'' (or XRD) radiative transfer.  
	
	\subsection{The effects of elastic collisions}
	
	\begin{figure*}[ht!]
		\centering
		{\includegraphics[scale=0.6]{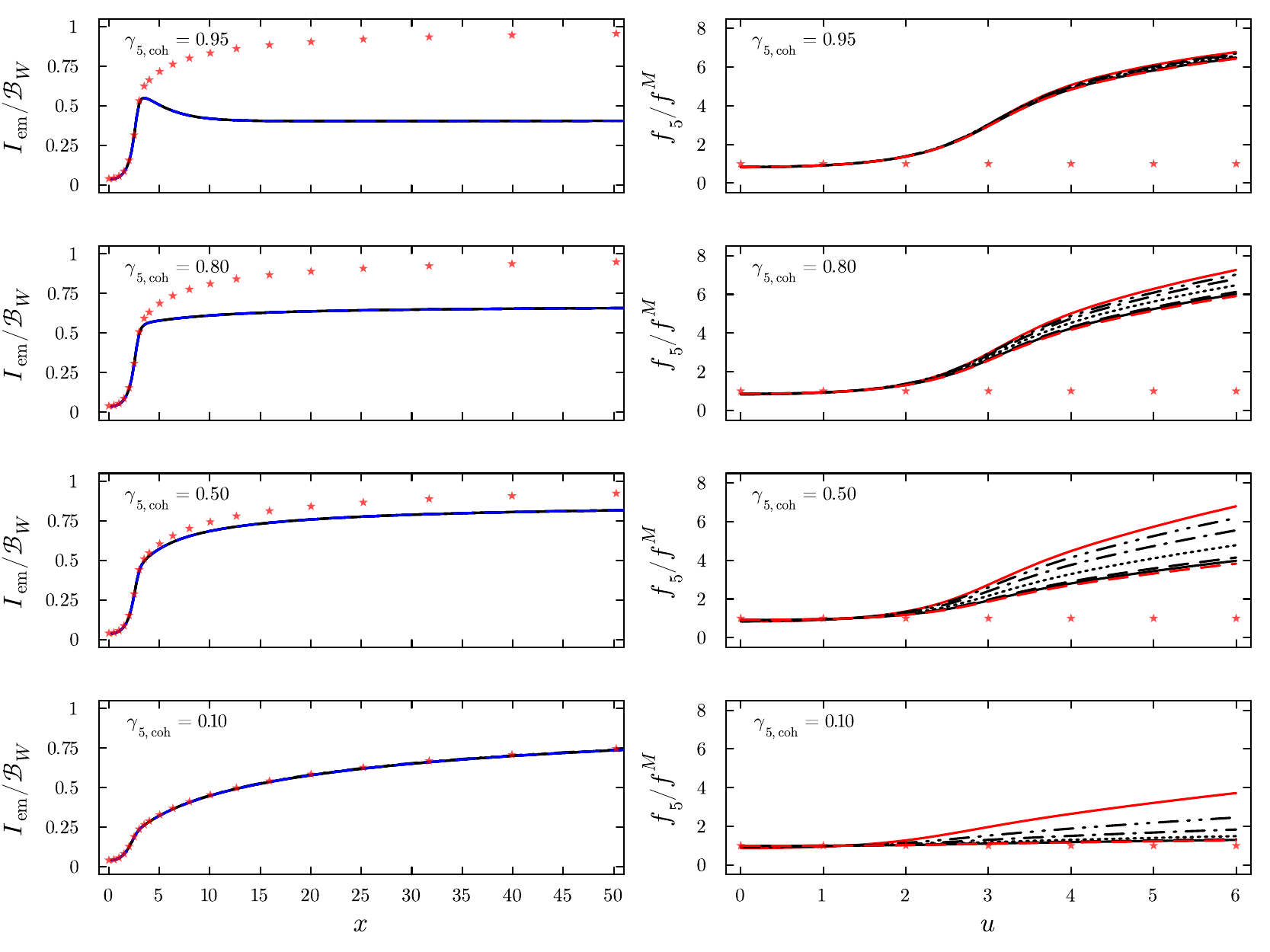}}
		\caption{Normalised emergent intensity at $\mu=1$ of the K line (left) and velocity distributions at $\tau=0$ of the fraction of atoms in level $5$ (right) shown for different values of the (total) collision rate $Q_E$ and its velocity-changing component $Q_V$. The amount of elastic collision increases from the top to the bottom with $\gamma_{5,\rm{coh}} = 0.95,\,0.80,\,0.50,\,0.10$. The dashed blue (red) line represents the emerging intensity (VDF) calculated with $Q_V=Q_E$ and the solid blue (red) line is calculated with $Q_V=0$. Between these two solutions, we represent $Q_V=0.9Q_E$ (solid black line), $Q_V=0.8Q_E$ (dashed black line), $Q_V=0.5Q_E$ (dotted black line), $Q_V=0.25Q_E$ (dash-dotted black line) and $Q_V=0.1Q_E$ (dash-double-dotted black line). For comparison, the red stars show the emergent intensity calculated in CRD (left) and the Maxwell-Boltzmann velocity distribution (right). }
		\label{fig:Fig5}
	\end{figure*}
	
	To accurately describe spectral line broadening, elastic collisions should be taken into account. Indeed, there are two processes that contribute to the total elastic collision rate: phase-changing and velocity-changing collisions. The total elastic collision rate is therefore the sum of the two rates $Q_E = Q_P + Q_V$. In this section, we study the effects of these collisions on the solutions to the radiation transfer problem and, by extension, on the formation of spectral lines. Apart from the total rate $Q_E$ of elastic collisions and $Q_V$ of v.c.c., all other parameters characterising our atmosphere remain identical. In Fig.~\ref{fig:Fig5}, we show the emergent intensity (left) of the K line of Ca\,{\sc ii} and the VDF associated with the fraction of atoms in level $5$ (right). The coloured dashed lines represent the solutions calculated with $Q_E=Q_V$ and the coloured solid lines represent those obtained with $Q_E=Q_P$. While it is seen that, for VDFs, there exists a set of intermediate solutions between these two solutions for varying amounts of v.c.c., the emergent intensity profiles are not affected by the inclusion of such collisions.  
	
	In order to analyse these results, we need to introduce a formulation of the problem in terms of redistribution functions which, although specific to two-level atoms, provide a relevant diagnostic tool for studying the solutions obtained in Fig.~\ref{fig:Fig5}. In the presence of v.c.c and for two-level atoms, it can be shown that the radiation transfer problem can be entirely reformulated through a single redistribution function defined as the linear combination of the pure CRD, i.e. $R^{\rm{CRD}}= \varphi(x')\varphi(x)$, and the redistribution function defined in Eq.~(\ref{eq:R}): 
	\begin{equation}
		R_{v.c.c}(x',x) = \gamma_V R(x',x) + (1 - \gamma_V)R^{\rm{CRD}}(x',x) \,,
	\end{equation} 
	with $\gamma_V = P_j/(P_j + Q_V)$ a branching ratio quantifying the importance of v.c.c in comparison with radiative and inelastic collisions. Defining also $\gamma_{\rm{coh}} = P_j / (P_j + Q_E)$, we can rewrite the branching ratio $\gamma_j$ as $\gamma_{\rm{coh}}/\gamma_V$. We then finally have:
	\begin{equation}
		\begin{split}
		R_{v.c.c}(x',x) &= \gamma_{\rm{coh}} R^{\rm{II}}(x',x) + (\gamma_V - \gamma_{\rm{coh}})R^{\rm{III}}(x',x) \\
		&+ (1 - \gamma_V)R^{\rm{CRD}}(x',x) \,.
		\end{split}
	\end{equation} 
	Although the redistribution function $R^{\rm{III}}$ is not well approximated in itself by the CRD redistribution function, we know that it leads to very similar emerging intensity profiles. Thus, we can write:
	\begin{equation}
			R_{v.c.c}(x',x) \approx \gamma_{\rm{coh}} R^{\rm{II}}(x',x) 
			+ (1 - \gamma_{\rm{coh}}) R^{\rm{CRD}}(x',x) \,.
	\end{equation} 
	Consequently, whether the v.c.c constitute all or part of the total amount of elastic collisions, \textit{only} the rate $Q_E$ of these collisions plays a direct role in the emerging intensity profiles, which explains the results shown in Fig.~\ref{fig:Fig5} (left panel). We also verify that the CRD solution can only be achieved in a highly collisional regime ($\gamma_{\rm{coh}},\gamma_V \rightarrow 0$). In such a regime, the lifetime of atomic levels is very short and the damping parameter becomes very large. Then as we see in Fig.~\ref{fig:Fig5}, the broadening of emergent intensity profile becomes more important as the elastic collisions rate increases. 
	
	On the other hand, we know that v.c.c appear in the kinetic equations of massive particle defined in Eq.~(\ref{eq:KEE}) via a relaxation (and approximate) term $n_iQ_V[f^M - f_i]$. Thus, when the v.c.c. are not significant in comparison with other collision processes, the deviations from Maxwellian become important (solid red lines).
	On the contrary, when the v.c.c are significant, the relaxation term becomes the dominant term of the KEEs and the velocity distribution tends to a Maxwellian, which can be clearly observed in Fig.~\ref{fig:Fig5} (red dashed lines). Strictly speaking, the CRD solution can only be achieved when all velocity distributions are Maxwellian. Even if VDFs are not Maxwellian in practice, only a small amount of elastic collisions is needed to lead to profiles close to those of the CRD.     
	
	While phase-changing collisions have already been studied by the community, to the best of our knowledge there are no studies on the influence of velocity-changing collisions, with the exception of the work by \citet[see in particular the discussion in Sect.~4, see also \citealt{Streaming3}]{SPV24}, and the present article. Here we show that the inclusion of v.c.c does not significantly affect the emerging intensity profiles, in the specific case of Ca\,{\sc ii}, for a semi-infinite atmosphere. However, the VDFs are strongly influenced by these collisions, suggesting that it is essential to estimate the v.c.c rate in future models.   
	
	\section{Discussions and conclusions}
	\label{sec:Discussion}

	In this article, we have developed a new numerical strategy for the solutions of the multi-level FNLTE radiative transfer problem. This method is based on approximate operator methods, which are widely used in the radiative transfer community. Our method consists in solving a MALI system for each of the atomic velocity in order to obtain, self-consistently, the velocity distributions associated with each atomic levels. For this reason, we have named this method MALIBU (Multilevel Accelerated Lambda Iteration, velocity $\vec{u}$ by velocity $\vec{u}$). Our new method is more robust than the one used by \citetalias{LPS25}. It is insensitive to the choice of initialisation and converges significantly faster. 
	
	Using the MALIBU method, we presented results for the FNLTE radiation transfer problem applied to a five-level model of Ca\,{\sc ii}. Under the physical conditions of the atmosphere studied and for this particular atomic model, we did not find any significant FNLTE effects on the emergent intensity profiles. However, we emphasise that we cannot prematurely generalise this conclusion. Indeed, in the case studied here, there is \textit{no} radiative transition between two lower levels. Consequently, the velocity distributions associated with these levels do not significantly deviate from the Maxwellian. However, we know that it is precisely these deviations that are mainly responsible for the deviations of the emission and absorption profiles from their equilibrium profiles. We also emphasise that a realistic modelling of the Ca\,{\sc ii} lines must take into account the photoionisation mechanism, as well as additional collisional and radiative transitions from higher energy levels. These physical processes tend to establish a natural population for the upper levels, thereby largely diminishing the effects of partial redistribution and cross redistribution in the standard NLTE as well as in our present FNLTE approach. It is possible to extend the current FNLTE approach to include these physical processes, although they are out of the scope of the present study. 
	
	In order to conclude definitively on the importance of FNLTE effects, we need to inspect the physical conditions leading to significant deviations of the VDFs associated with the lower levels and study increasingly complex atomic models for which there are radiative transitions between two excited levels. A good candidate is then hydrogen (or sodium), for which, however, frequency redistribution must be treated very carefully \citep{HubenyLites1995}. It is also possible to investigate other atmospheric conditions, such as strongly illuminated finite slabs, for which greater deviations from the Maxwellian velocity distribution are expected as shown by \citet{PSP23}.
	
	We also discussed the effects of elastic collisions on the spectral line formation. There are two components to these collisions, one changing the phase and the other changing the velocity of massive particles (see e.g. \citealt{SPV24} for more details). We show that these collisions do not significantly alter the emerging intensity profiles. This suggests that, under the conditions studied, the standard NLTE formalism with partial redistribution in frequency is sufficient to understand the emerging intensity profiles associated with the Ca\,{\sc ii} spectral lines. However, these collisions have a significant effect on the calculation of velocity distributions. Only our approach allows such quantities to be calculated. It is therefore relevant to accurately quantify the rate of velocity-changing collisions (see e.g. \citealt{SPV24}).               
	
	More generally, our present model still contains several approximations. Some simplifying assumptions, like the isotropy of VDFs, still need to be evaluated in practice. Other assumptions are much more structural and require us to review all or part of our formalism. This is the case for the streaming of massive particles or stimulated emission. We can consider including the effects of streaming after the work of \cite{Streaming1,Streaming2} and \cite{Streaming3}. Taking stimulated emission into account is much more delicate, as pointed out by \cite{HOSI}. Several approaches can be considered. One developed by \cite{OxeniusSimoneau94} within the framework of the Weisskopf-Woolley model, solving a set of kinetic equations acting directly on the population of atomic sublevels. Another proposed by \cite{HOSI}, where all the atomic profiles are obtained consistently using an iterative procedure. In any case, improving our FNLTE approach will require a revision of some of the numerical strategies and tools used until now.   
	
	\begin{acknowledgements}
		T. Lagache is kindly supported by a ``PhD booster'' 
                grant by the Toulouse Graduate School of Earth and
                Space Sciences (TESS; \url{https://tess.omp.eu/}). The authors would like to thank the reviewer, Prof. Ivan Huben{\'y}, for his very valuable, critical and insightful comments. 
	\end{acknowledgements}
	
	\bibliographystyle{aa}
	\bibliography{aa60219-26_biblio.bib}
	
	\begin{appendix}
		
		\section{Numerical evaluation of absorption profiles and generalised redistribution functions}
		
		\begin{figure}[ht!]
			\includegraphics[width=\columnwidth]{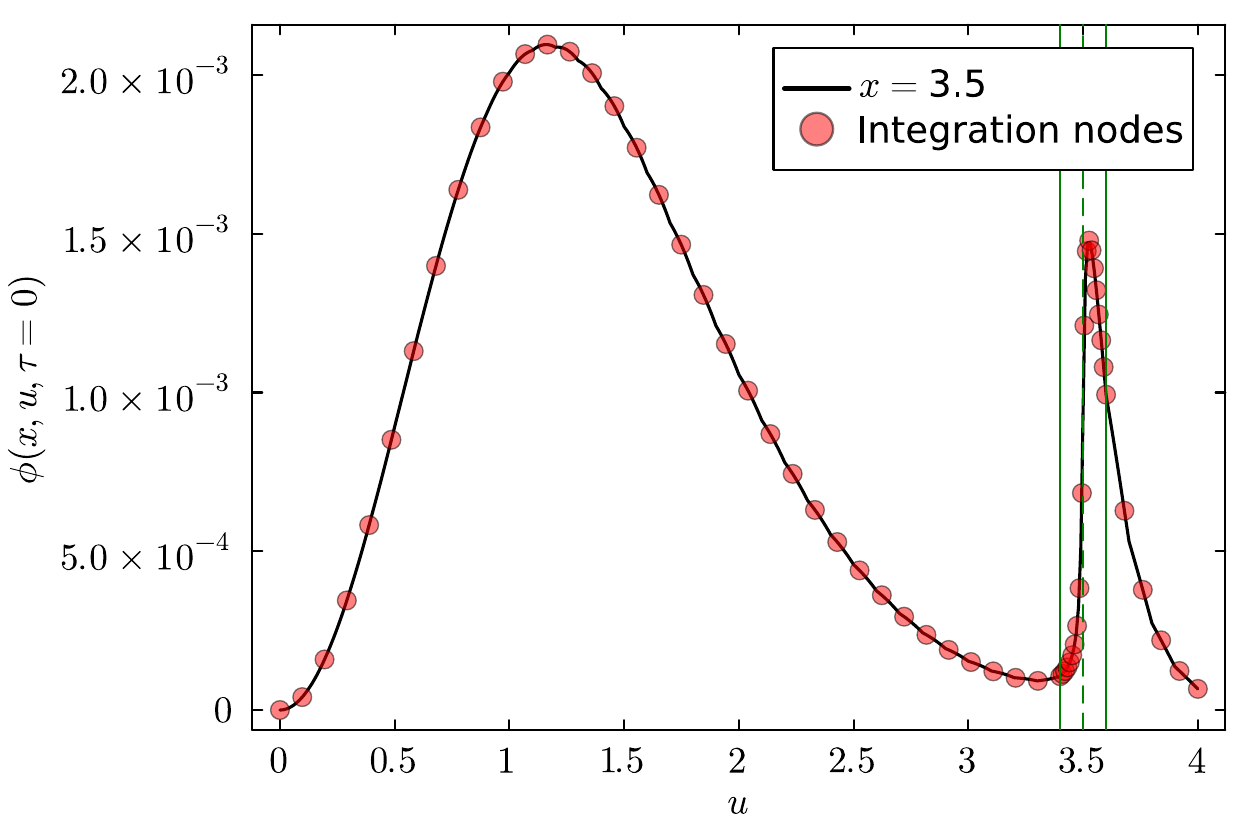}
			\caption{Integrand $\phi$ computed with the velocity distribution of the first excited level from \cite{SPV24}, for a damping parameter $a=10^{-2}$. Its variation in normalised velocity $u$ is shown for $x=3.5$ and $\tau=0$ (black line). The red dots represent the integration nodes used to calculate the profile $\varphi(x,\tau)$ defined in Eq.~(\ref{eq:phifp}). Between the two vertical green lines, the density of the integration nodes is greater to take into account the peak located at $u=\lvert x \lvert$.}
			\label{fig:AnnexeAFig1}
		\end{figure} 
		
		We propose to numerically evaluate the absorption profiles and generalised redistribution functions (Eqs~.\ref{eq:phifl},\ref{eq:RIII},\ref{eq:RII}), calculated with a priori non-Maxwellian VDFs. The methods used for these two quantities are very similar, and we will only present here the method used to calculate the absorption profiles. We therefore aim to numerically evaluate the following quantity: 
		\begin{equation}
			\begin{split}
			\varphi^{p*} &= \int_0^\infty \frac{uf_p(u)}{2\pi} \left[\arctan \left(\frac{u+x}{a}\right) + \arctan \left(\frac{u-x}{a}\right) \right] du \\
			&= \int_0^\infty \phi(x,u,\tau) du \,,
			\end{split} 
			\label{eq:phifp}
		\end{equation}
		First, we know that $\forall u \geq 0$:
		\begin{equation}
			\left[\arctan \left(\frac{u+x}{a}\right) + \arctan \left(\frac{u-x}{a}\right) \right] \xrightarrow[ a \rightarrow 0]{}  \pi H(u-\lvert x \lvert) \,,
		\end{equation}
		where $H$ is the Heaviside function. Consequently, the integrand $\phi$ will be more peaked around $u=\lvert x \lvert$ when the damping parameter $a$ is small. At each depth and for each frequency, we must integrate the function $\phi$ over all atomic velocities. This calculation needs to be repeated several times, so it is necessary to implement an integration strategy that is both accurate and fast. We will use a simple trapezoidal quadrature. Beyond a certain limit, typically for $\lvert x \lvert >u_{\rm{max}}+10a$, we estimate that the peak at $u = \lvert x \lvert$ is negligible\footnote{For $u_{\rm{max}} > 6$, we have $f_p(u) \propto e^{-u_{\rm{max}} ^2} \ll 1$.}. Thus, in such a situation, the integration grid will simply be the velocity grid. However, when $\lvert x \lvert < u_{\rm{max}}+10a$, it is important to resolve this peak accurately. We then subdivide the grid into two parts. The first part is centred on the peak and going from $u^-=\lvert x \lvert - 10a$ to $u^+= \lvert x \lvert +10a$. In this range, we regularly space $20$ integration nodes, with a step size of $du=a$. Then a second part such that $u$ is contained in $\left[0, u^-\right[\, \cup \, \left]u^+,u_{\rm{max}}\right]$, with $N>3$ points spaced regularly with $du\approx0.1$. An example is shown in Fig~.\ref{fig:AnnexeAFig1} for $\tau=0$, $x=3.5$ and $a=10^{-2}$. The VDF used here is the one calculated by \cite{SPV24} for a two-level hydrogen atom with natural broadening of the upper level.  
		
		\begin{figure}[ht!]
			\includegraphics[width=\columnwidth,]{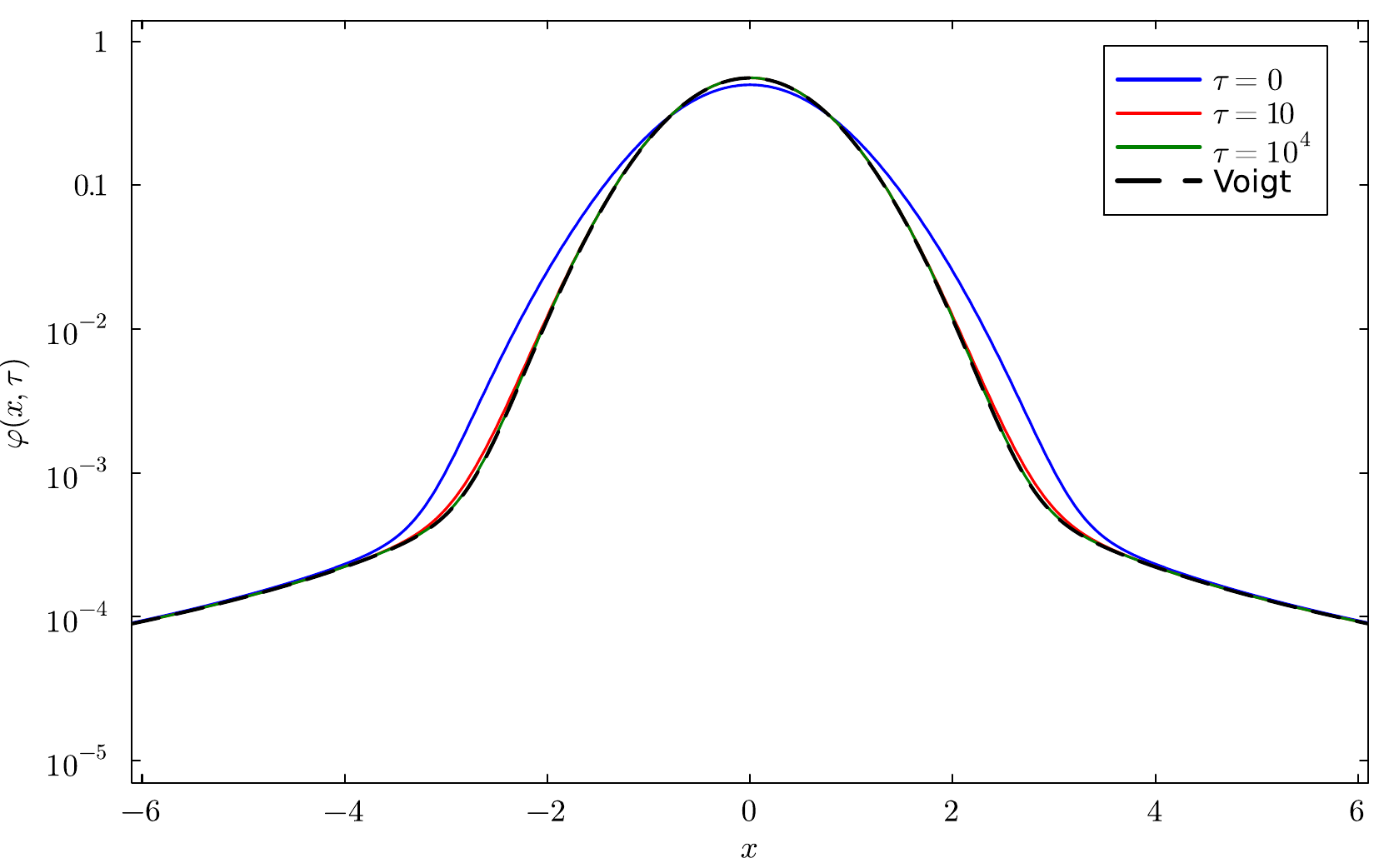}
			\caption{Absorption profile $\varphi$ calculated using the velocity distribution function of the first excited level from \cite{SPV24} and a damping parameter $a=10^{-3}$. Its variation in frequency is shown for different optical depths $\tau=0,10,10^4$ (coloured lines). As the depth increases, the velocity distribution becomes Maxwellian and the absorption profile approaches the Voigt profile (black dashed line).}
			\label{fig:AnnexeAFig2}
		\end{figure}  
		
		Using the same velocity distribution, Fig.\ref{fig:AnnexeAFig2} shows the absorption profile calculated with $a=10^{-3}$. As the optical depth $\tau$ increases, the VDF approaches the Maxwellian $f^M$ and thus, the absorption profile approaches the Voigt profile. When $f^p = f^M$, our method allows us to numerically compute the Voigt profile with an average accuracy of $0.8\%$ in comparison with the Voigt-Fadeeva function, which is very satisfactory in comparison with other methods such as that of \cite{AGQ}, which obtains an error of the same order\footnote{The method of \cite{AGQ} is, however, more accurate for $a\lesssim 10^{-3}$ but less so for $a\gtrsim 10^{-3}$.}. 
		
		\begin{figure}[t!]
			\includegraphics[width=\columnwidth,]{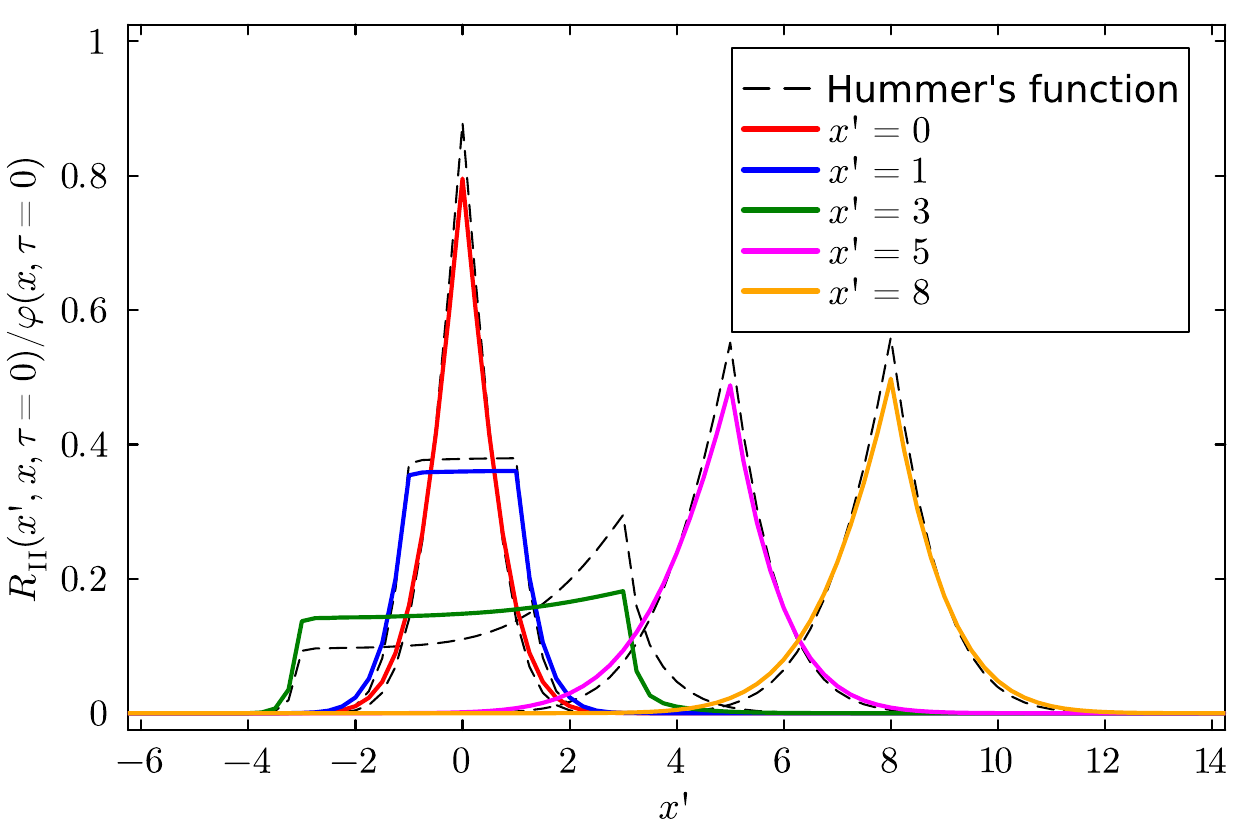}
			\caption{Frequency variation $x'$ of the ratio of the correlated generalised redistribution function $R_{\rm{II}}(x',x,\tau=0)$ and the absorption profile $\varphi(x,\tau=0)$ for different frequencies $x=0,1,3,5,8$ (coloured lines) and calculated with the velocity distribution of the first excited level from \cite{SPV24}. We have $a=10^{-3}$ and $\alpha=1$. The black dashed lines show the same ratio calculated with a Maxwellian distribution, which corresponds to the redistribution function from \cite{hummer62_redistrib}. }
			\label{fig:AnnexeAFig3}
		\end{figure} 
		
		\begin{figure}[t!]
			\includegraphics[width=\columnwidth,]{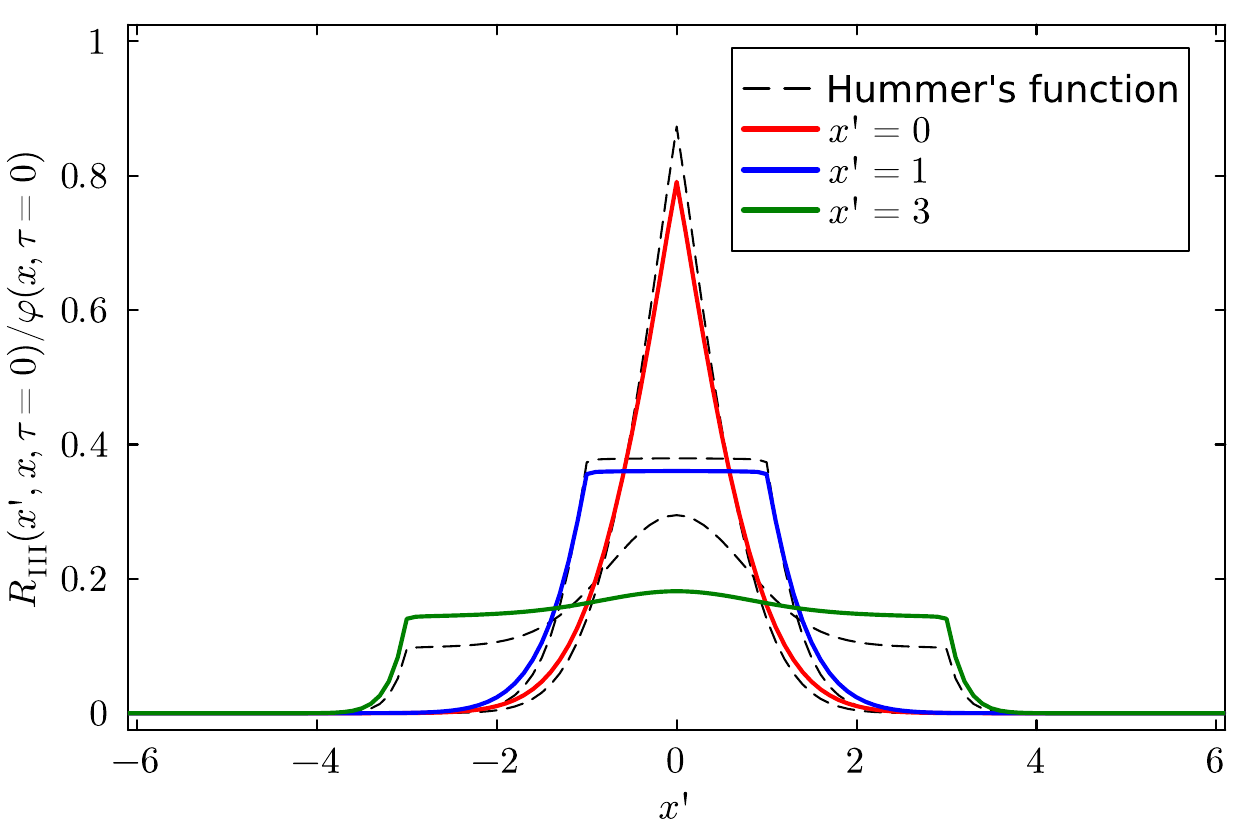}
			\caption{Same as Fig.~\ref{fig:AnnexeAFig3} but for the decorrelated redistribution function. We took $a=a'=10^{-3}$.}
			\label{fig:AnnexeAFig4}
		\end{figure} 
		
		A similar method can be used to evaluate generalised redistribution functions. Figures~\ref{fig:AnnexeAFig3} and\,\ref{fig:AnnexeAFig4} show the ratio between the redistribution functions and the absorption profile, calculated using the same VDF as before. We only show the results obtained for a redistribution mechanism occurring within the same line. It is clear that our results (coloured lines) differ significantly from the results of \citet[dashed lines]{hummer62_redistrib}, which are found deep in the atmosphere. This shows the need to compute the velocity distribution of massive particles self-consistently in order to accurately understand the mechanisms of frequency redistribution.   
		
		\label{app:AppA}

	\end{appendix}
	
\end{document}